\documentclass[aps,preprint,superscriptaddress]{revtex4-2}
\usepackage[pdftex]{graphicx}
\usepackage{setspace,ragged2e}
\usepackage[monochrome]{color}  
\usepackage[letterpaper, total={7.28in, 9.55in}]{geometry}
\usepackage{physics}
\usepackage{hyperref}
\hypersetup{
    colorlinks=true,
    linkcolor=blue,
    filecolor=blue,      
    urlcolor=blue,
    citecolor=blue
}
\usepackage{pdfpages}
\makeatletter 
\AtBeginDocument{\let\LS@rot\@undefined}
\makeatother
\usepackage{pgffor}
\newcommand{\cpd}{CePd$_3$}
\begin{document}

\title{Kondo quasiparticle dynamics observed by\\resonant inelastic x-ray scattering}

\author{M. C. Rahn}
\email[]{marein.rahn@tu-dresden.de}
\affiliation{Los Alamos National Laboratory, Los Alamos, New Mexico 87545, USA}
\affiliation{Institute for Solid State and Materials Physics, Technical University of Dresden, 01062 Dresden, Germany}

\author{K. Kummer}
\affiliation{European Synchrotron Radiation Facility, BP 220, F-38043 Grenoble Cedex, France}

\author{A. Hariki}
\affiliation{Department of Physics and Electronics, Graduate School of Engineering, Osaka Prefecture University, Sakai, Osaka 599-8531, Japan}
\affiliation{Institute for Solid State Physics, TU Wien, 1040 Vienna, Austria}

\author{K.-H. Ahn}
\affiliation{Institute for Solid State Physics, TU Wien, 1040 Vienna, Austria}
\affiliation{Institute of Physics of the CAS, Cukrovarnick\'{a} 10, 162 00 Praha 6, Czechia}

\author{J. Kune\v{s}}
\affiliation{Institute for Solid State Physics, TU Wien, 1040 Vienna, Austria}

\author{A. Amorese}
\affiliation{Institute of Physics II, University of Cologne, Cologne, Germany}
\affiliation{Max Planck Institute for Chemical Physics of Solids, Dresden, Germany}

\author{J. D. Denlinger}
\affiliation{Advanced Light Source, Lawrence Berkeley Laboratory, Berkeley, California 94720, USA}

\author{D.-H. Lu}
\affiliation{Stanford Synchrotron Radiation Lightsource, SLAC National Accelerator Laboratory, Menlo Park, California 94025, USA}

\author{M. Hashimoto}
\affiliation{Stanford Synchrotron Radiation Lightsource, SLAC National Accelerator Laboratory, Menlo Park, California 94025, USA}

\author{E. Rienks}
\affiliation{Helmholtz Zentrum Berlin, Bessy II, D-12489 Berlin, Germany}

\author{M. Valvidares}
\affiliation{ALBA Synchrotron Light Source, E-08290 Cerdanyola del Vall\`{e}s, Barcelona, Spain}

\author{F. Haslbeck}
\affiliation{Physik-Department, Technische Universität München, D-85748 Garching, Germany}
\affiliation{Institute for Advanced Studies, Technische Universität München, D-85748 Garching, Germany}

\author{D. D. Byler}
\affiliation{Los Alamos National Laboratory, Los Alamos, New Mexico 87545, USA}

\author{K. J. McClellan}
\affiliation{Los Alamos National Laboratory, Los Alamos, New Mexico 87545, USA}

\author{E.~D. Bauer}
\affiliation{Los Alamos National Laboratory, Los Alamos, New Mexico 87545, USA}

\author{J.-X. Zhu}
\affiliation{Los Alamos National Laboratory, Los Alamos, New Mexico 87545, USA}

\author{C.~H. Booth}
\affiliation{Chemical Sciences Division, Lawrence Berkeley National Laboratory, Berkeley, California 94720, USA}

\author{A.~D. Christianson}
\affiliation{Oak Ridge National Laboratory, Oak Ridge, Tennessee 37831, USA}

\author{J. M. Lawrence}
\affiliation{Department of Physics and Astronomy, University of California, Irvine, California 92697, USA}
\affiliation{Los Alamos National Laboratory, Los Alamos, New Mexico 87545, USA}

\author{F. Ronning}
\affiliation{Los Alamos National Laboratory, Los Alamos, New Mexico 87545, USA}

\author{M. Janoschek}
\email[]{marc.janoschek@psi.ch}
\affiliation{Los Alamos National Laboratory, Los Alamos, New Mexico 87545, USA}
\affiliation{Institute for Advanced Studies, Technische Universität München, D-85748 Garching, Germany}
\affiliation{Laboratory for Neutron and Muon Instrumentation, Paul Scherrer Institute, CH-5232 Villigen, Switzerland}
\affiliation{Physik-Institut, Universit\"{a}t Z\"{u}rich, Winterthurerstrasse 190, CH-8057 Z\"{u}rich, Switzerland}


\maketitle

\textbf{Effective models focused on pertinent low-energy degrees of freedom have substantially contributed to our qualitative understanding of quantum materials. An iconic example, the Kondo model, was key to demonstrating that the rich phase diagrams of correlated metals originate from the interplay of localized and itinerant electrons. Modern electronic structure calculations suggest that to achieve quantitative material-specific models, accurate consideration of the crystal field and spin-orbit interactions is imperative. This poses the question of how local high-energy degrees of freedom become incorporated into a collective electronic state. Here, we use resonant inelastic x-ray scattering (RIXS) on CePd$_3$ to clarify the fate of all relevant energy scales. We find that even spin-orbit excited states acquire pronounced momentum-dependence at low temperature ---the telltale sign of hybridization with the underlying metallic state. Our results demonstrate how localized electronic degrees of freedom endow correlated metals with new properties, which is critical for a microscopic understanding of superconducting, electronic nematic, and topological states.}

\begin{figure*}
\includegraphics[width=1\columnwidth,trim= 0pt 0pt 0pt 0pt, clip]{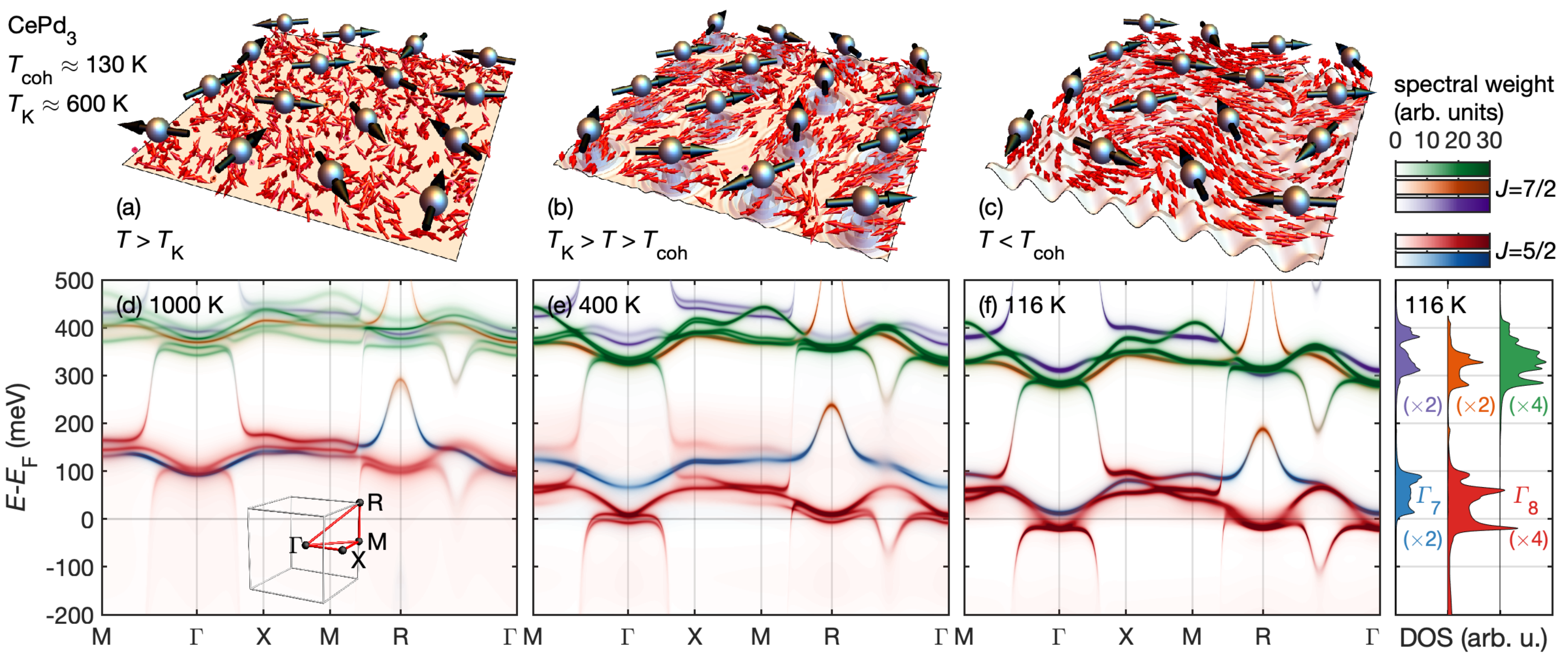}%
\justifying
\caption{\label{Fig1}  \textbf{Emergence of a coherent Kondo lattice state as a function of temperature in \cpd.} (a-c) Real-space impressions of the increasingly coherent screening of Ce moments (black arrows) across the temperature scales $T_\mathrm{K}$ and $T_\mathrm{coh}$. Spins of itinerant electrons are represented by red arrows (see text for details). (d-f) Electronic spectral function $A(k,\omega)$ of \cpd~obtained from the combination of density functional and dynamical mean field theory (DFT+DMFT) at 1000\,K, 400\,K, and 116\,K. The fourteen $4f$ orbitals of cerium form seven Kramers doublets, separated into a $J=5/2$ sextuplet and a $J=7/2$ octuplet by $E_\mathrm{SO}\approx280$\,meV. A cubic crystal field (CF) further induces a $E_\mathrm{SO}\approx10$\,meV splitting between a $J=5/2$ $\Gamma_7$ doublet (blue) and the $\Gamma_8$ quartet (red). The $J=7/2$ states are split into two doublets (purple and orange) and one quartet (green). (a,d) Above the Kondo temperature, $T_\textrm{K}\approx600$\,K, local $f$-electron moments are uncorrelated. The corresponding electronic structure only shows weak hybridization with the incoherent $f$-states. (b,e) For $T<T_\textrm{K}$, local $f$-electron magnetic moments effectively couple to the surrounding conduction electrons, forming virtual-bound states in the vicinity of the Fermi energy $E_F$ (suggested as ripples in Panel (b)). The $f$ spectral weight close to the Fermi surface remains predominantly incoherent. (c,f) Below the coherence temperature $T_\text{coh}\approx130$\,K, the $f$-states are coherently incorporated (suggested by the lattice-coherent modulation in Panel (c)) into the underlying band structure as dispersive, albeit heavy, electronic quasiparticles. The right panel shows the  density of states (DOS) summed over the displayed high-symmetry directions, projected onto the different CF characters.}
\end{figure*}

Metals containing lanthanides are optimally suited to study the Kondo lattice problem, owing to highly localized $4f$-electrons, which at low temperature form a strongly-correlated many-body state~\cite{Paschen2021,Lawrence2008}. The Kondo lattice model is characterized by two effective low-energy scales~\cite{Burdin2000} (see Fig.~\ref{Fig1}(a-c)): the Kondo temperature $T_\textrm{K}$, below which local $f$-electron magnetic moments become effectively coupled to the surrounding conduction electrons, and the coherence temperature $T_\text{coh}$, below which these states are coherently incorporated into the underlying band structure as dispersive, albeit heavy, electronic quasiparticles. A large body of experiments has shown that real materials can indeed be qualitatively characterized by these two energy scales. Modern angle-resolved photoemission spectroscopy (ARPES)~\cite{Jang2017}, quasiparticle interference~\cite{Aynajian2012}, and inelastic neutron scattering (INS)~\cite{Goremychkin2018} have even directly observed the formation of coherent heavy quasiparticle bands. Frustrated magnetism~\cite{Fobes2018}, unconventional superconductivity~\cite{Pfleiderer2009}, hidden order~\cite{Mydosh2020}, as well as topologically non-trivial~\cite{Pirie2020, Kurumaji2019, Jiao2020} and electronic-nematic states~\cite{Ronning2017, Rosa2020} are all known to emerge from this Kondo lattice state. 

The two phenomenological energy scales ($T_\textrm{K}$, $T_\text{coh}$) clearly do not suffice to account for this variety of experimental discoveries. Instead, material-specific knowledge of the quasiparticle fine structure will be required to disentangle the underlying interaction channels and achieve a guided design of emergent functional properties. Magnetic, thermodynamic~\cite{Pagliuso2002}, and x-ray spectroscopic~\cite{Willers2015} studies of crystal-field (CF) ground states in Kondo lattices have indeed demonstrated that the anisotropic hybridization of the $f$-electron wave function strongly influences the ground state properties, even for materials with the same crystal structure. Moreover, bulk measurements as a function of external tuning parameters like magnetotransport and magnetostriction suggest that the structure of unoccupied crystal field levels cannot be disregarded in these investigations~\cite{Moll2017, Rosa2019}. Realistic models of these circumstances can be achieved by the combination of density functional and dynamical mean field theory (DFT+DMFT). And indeed, such calculations have recently confirmed that high energy scales like CF and SO splitting must be taken into account to quantitatively reproduce low energy ground states~\cite{Shim2007,Haule2010, Jang2017}.

Clarifying the mechanisms by which the properties of localized $f$-electrons blend into an itinerant environment is an outstanding experimental challenge. While local probes capture the overall energy scales of CF and SO excited states, they cannot observe their momentum dependence---the telltale sign of hybridization with the underlying metallic state. This crucial microscopic knowledge requires low-energy momentum-resolved spectroscopic methods like quasiparticle interference or INS. Recent pioneering ARPES studies~\cite{Jang2017,Patil2016,Kummer2015} have directly mapped anisotropic hybridization close to the Fermi energy in Ce$M$In$_5$ and Ce$M_2X_2$ heavy fermion materials ($M=$ transition metal; $X=$ Si, Ge). This new level of material-specific microscopic insights has shone a light on the relevance of crystal field and magnetic degrees of freedom, and has called into question widely accepted concepts like the direct relation between the temperature scale $T_\mathrm{coh}$ and the formation a ``large'' Fermi surface~\cite{Kummer2015}. One common constraint of these techniques is that they lack sensitivity to the unoccupied $f$-electronic structure at higher energies~\cite{Murani1996b}. To overcome this limitation, we take advantage of instrumental advances that have made resonant inelastic x-ray scattering (RIXS) a powerful probe of Kondo lattices~\cite{Hancock2018,Amorese2018c}.

$M$-edge RIXS is a photon-in/photon-out process in which $3d$ core electrons are resonantly excited and recombine by  dipole-transitions after interacting with $4f$ valence states. Like INS, this serves as a probe of two-particle correlations, providing a complementary perspective on the single-particle spectral function measured by ARPES. It has the advantage of true bulk-sensitivity at good energy resolution, but, in contrast to INS, the observation of inter-/intraband transitions by x-rays is not restricted to the spin-flip channel. RIXS also has major experimental advantages, as it is not dominated by phonon scattering, and does not require large single crystals and long counting times. Crucially for the present context, the insights are neither limited by the instrumental energy resolution, nor in the range of energy transfer, and the resonant character and polarization dependence allow to separate different excitation channels~\cite{Amorese2018c}.

We focus on the archetypal Kondo lattice material \cpd, with an effective $f$-occupation of $n_f\approx0.75$ (from x-ray absorption spectroscopy~\cite{Fanelli2014}), and $k_\mathrm{B} T_\mathrm{K}\approx55$\,meV. The Kondo energy of this system is thus intermediate to the bare SO coupling $E_\mathrm{SO}\approx280$\,meV of Ce and the coherence energy scale  $k_B T_\text{coh}\approx11$\,meV (derived from transport data)~\cite{Lawrence1985} and CF splitting $E_\mathrm{CF}\approx10$\,meV (from comparison to the isostructural material CeIn$_3$~\cite{Knafo2003} and the present DFT calculation). This cascade of energy scales makes \cpd~ideally suited to address both our experimental and computational objectives, i.e., to understand how crystal field and spin-orbit interactions become coherently incorporated into the Kondo lattice state. The relatively large $T_\mathrm{coh}$ and $T_\mathrm{K}$ ensure that temperatures accessible by DFT+DMFT are sufficient to simulate coherence~\cite{Goremychkin2018}, and the quasiparticle bandwidth is large compared to the energy resolution $\mathrm{d}E\sim35$\,meV of RIXS.

Fig.~\ref{Fig1} shows the correlated band structure of \cpd~at 1000\,K, 400\,K, and 116\,K, as obtained via DFT+DMFT calculations to visualize the emergence of a coherent Kondo lattice in \cpd. Cerium features fourteen $4f$ orbitals corresponding to seven Kramers doublets, separated into a $J=5/2$ sextuplet and a $J=7/2$ octuplet by $E_\mathrm{SO}$. The cubic CF further splits the $J=5/2$ and $J=7/2$ states into a total of five CF multiplets indicated in Fig.~\ref{Fig1} in distinct colors. Fig.~\ref{Fig1}(d) illustrates that even at $T$~$=$~1000\,K~$>$~$T_\textrm{K}$, the  modulation of the $4f$ states due to the onset of anisotropic hybridization outweighs the weak CF splitting, which underscores the importance of momentum-resolved data. Below $T_\textrm{K}\approx600$\,K, the increasing hybridization renormalizes all $4f$-states. However, their spectral weight close to the Fermi surface remains predominantly incoherent (Fig.~\ref{Fig1}(e)). For $T<T_\textrm{coh}\approx130$\,K, coherence sets in, and well-defined quasiparticle bands that cross the Fermi level form pockets at the $\Gamma$ and R points (Fig.~\ref{Fig1}(f)). Our calculations emphasize the amount of information encoded in the unoccupied states. Instead of the $\Gamma_7$ CF ground state of isostructural (and more localized) CeIn$_3$~\cite{Thalmeier1984}, we find that the hybridization with Pd $d$ bands endows the quasiparticle pockets with $\Gamma_8$-like character akin to CeB$_6$~\cite{Sundermann2017} (for CePd$_3$ at 116\,K we obtain occupation numbers $n_{\Gamma7}=0.08$ and $n_{\Gamma8}=0.68$). The SO excited states are similarly modified due to hybridization. Fig.~\ref{Fig1}(f) shows the summed density of states of different orbital characters. We find that the $J=7/2$-like states retain an occupation number of 0.17, even at low temperatures (116 K). This indicates that, aside from the mixing of CF states, strictly speaking, even $J$ cannot be considered a good quantum number ($E_\mathrm{K}\approx 0.2\,E_\mathrm{SO}$).

Recently, Goremychkin et al.~\cite{Goremychkin2018} used state-of-the-art DFT+DMFT calculations, like those presented here, to quantitatively model their neutron spectra of magnetic inter-band transitions and conclusively demonstrate the formation of coherent quasiparticle bands in \cpd. This pioneering work has raised several remaining key issues on the nature of the coherent state: (i) To date, the phenomenological temperature scale $T_\text{coh}$ has only been inferred from bulk properties like maxima in resistivity and magnetic susceptibility~\cite{Lawrence1985} and the onset of phase-coherence remains to be explored microscopically. (ii) The INS study~\cite{Goremychkin2018} was limited to energy transfers between 20 and 65\,meV, but also predicted excitations within the putative SO gap, which calls into question how coherence impacts the coupling of spin and orbital momentum. (iii) To clarify the fate of CF ground states assigned to heavier Ce systems, it would be crucial to resolve the fine structure of the quasiparticle bands.

To address these challenges, our study has to overcome the limitations of INS as a probe of electronic structure and encompass all energy scales relevant for DFT+DMFT. To complement RIXS, we present first single-crystal ARPES data of \cpd, resolving electronic structure within few meV of the Fermi level. Moreover, we corroborate the two-particle dynamics in the 50$\sim$250\,meV regime by higher-energy INS (see Section 3 of the Supplementary Information (SI)~\cite{supplemental} file), and use x-ray absorption spectroscopy (XAS) to probe the RIXS intermediate states in the presence of a Ce 3$d_{5/2}$ core hole. This provides fixed points for a global, realistic model of the inter-band transitions.

\begin{figure*}
\includegraphics[width=1\columnwidth,trim= 0pt 0pt 0pt 0pt, clip]{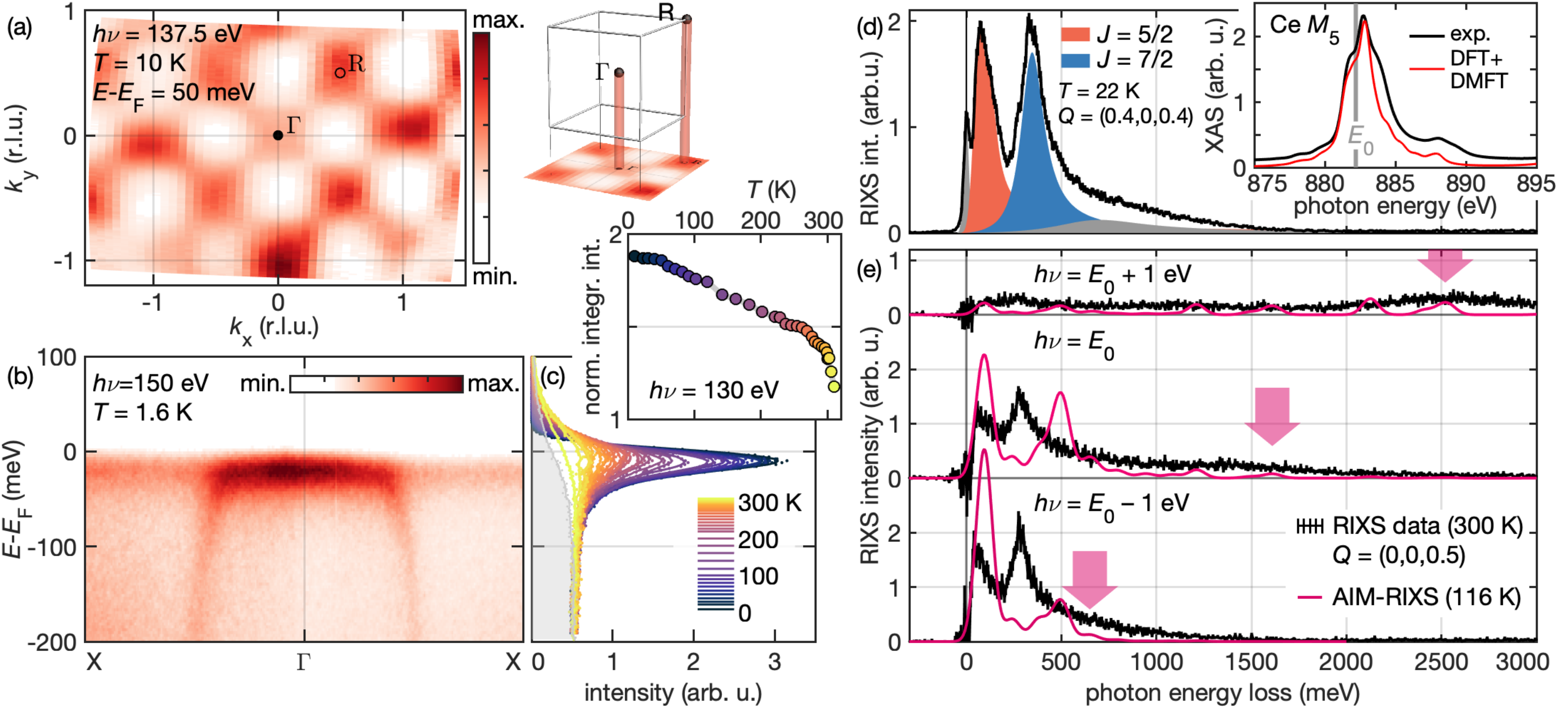}%
\caption{\label{Fig2} \textbf{Comparison of our model with spectroscopic data, covering the electronic structure of \cpd~from the meV to the eV scale.} For a global comparison of the DFT+DMFT calculation with the prototypical Kondo lattice material, we employ three spectroscopic methods. (a-c) Vacuum ultraviolet angle-resolved photo emission spectroscopy (ARPES) reveals the the electronic spectral function $A(k,\omega)$ with meV resolution. (a) Low-temperature quasiparticle Fermi surface observed by ARPES. As the momentum along the surface normal is not resolved, the pockets $\Gamma$ and R are projected onto the $k_x$--$k_y$ plane (cf. illustration of the Brillouin zone). (b) High-resolution ARPES spectrum highlighting the hybridization of the almost vertical Pd $4d$ bands with the Ce $4f$ states at the $\Gamma$ point. (c) Temperature dependence of energy distribution curves of this Kondo resonance. The gray shaded area is the reference spectrum of a gold foil at 310\,K. The inset shows the integrated intensity of $f$-electronic density of states normalized to the integrated intensity of the metallic background, as approximated by the gold spectrum. (d,e) Resonant inelastic x-ray scattering (RIXS) covering energy transfers of several eV. (d) Overview of the characteristic features of CePd$_3$ RIXS at $E_0=882.2$\,eV, here shown for $Q=(0.4,0,0.4)$, at 22\,K. Spectral weight assigned to excitations of either spin orbit state is marked by colored Lorentzian lineshapes. The resolution-limited elastic peak and a sloping background due to excitations into the continuum of Pd $4d$ states are shaded gray. The inset shows a comparison of the low-temperature Ce M$_5$ X-ray absorption spectroscopy (XAS) spectrum with our DFT+DMFT calculation. (e) Comparison of a $Q=(0,0,0.5)$ RIXS spectrum at $E_0$ with data recorded 1\,eV above/below this resonance (elastic line subtracted). Pink lines show the respective results of the AIM-RIXS calculation at 116 K. Aside from the Raman-like $J=5/2$  and $J=7/2$  peaks, the calculation captures the shift of continuum excitations to higher energy transfers at higher photon energies $h\nu$ (indicated by arrows). Due to the necessarily coarse discretization, the SO gap is consistently overestimated and the continuum is mapped onto a series of peaks.}
\end{figure*}

Fig.~\ref{Fig2}(a,b) shows the occupied quasiparticle states in the vicinity of the Fermi level of a $[1,0,0]$-fractured sample at low temperatures obtained via single-crystal vacuum ultraviolet ARPES measurements. As predicted by DFT+DMFT, quasiparticle Fermi pockets form at the $\Gamma$ and R points. Since the electron momentum perpendicular to the sample surface ($k_z$) is not resolved in these measurements, both $\Gamma$ and R hot spots are projected onto the $k_x$-$k_y$ plane. The observed characteristic checkerboard-pattern of the dispersion around the $\Gamma$ point confirm that DFT+DMFT yields an adequate model of the coherent Kondo lattice state with regard to the quasiparticles below the Fermi energy. Fig.~\ref{Fig2}(c) illustrates how spectral weight due to increasing hybridization of $f$-states continuously rises at low temperatures and forms a $\sim30$\,meV wide Kondo peak at the Fermi surface, which is closely reminiscent of Fig.~\ref{Fig1}(d--f). 

Now we turn to discuss our RIXS measurements. Fig.~\ref{Fig2}(d) shows a typical M$_5$ RIXS spectrum of \cpd, recorded at 22\,K (i.e., well below the coherence temperature), at a photon energy of $E_0=882.2$\,eV, where excitations of $f$-quasiparticles with $J=5/2$ and $J=7/2$-like character carry similar spectral weight (as indicated by phenomenological colored Lorentzian line shapes). The two SO bands are much wider than the instrumental energy resolution, which can be inferred from the elastic line, a Gaussian peak of 32\,meV full-width at half-maximum. Their intensity appears in the crossed ($\sigma\pi'$) polarization channel and is likely of magnetic character (see Supplementary Note 6~\cite{supplemental}). We find no additional two-particle excitation in the $\sim200$\,meV regime. Instead, up to this energy, RIXS is in quantitative agreement with the dynamic magnetic susceptibility $\chi''(Q,\omega)$ probed by INS (see our detailed discussion in Supplementary Note 3~\cite{supplemental}). However, neutron excitations across the spin-orbit gap are suppressed by a factor $\sim200$ ~\cite{Murani1996b}, which makes it de facto impossible to probe momentum dependence of excited $f$-electron states with INS. As a combination of two consecutive dipole transitions, RIXS breaks free of these constraints. Moreover, by selecting the intermediate state and using polarization analysis, it is possible to separate the $J=5/2$ and $J=7/2$ RIXS spectral weight~\cite{supplemental}.

For a realistic model of RIXS in \cpd~going beyond the comparison to INS, we compute the Kramers-Heisenberg cross section in an Anderson impurity model based on the DFT+DMFT electronic structure~\cite{supplemental}. Notably, this AIM-RIXS calculation relies on a single parameter, the double-counting correction $\mu_{dc}$, which shifts the energy of the $4f$ states to take into account the $f-f$ interaction present in the DFT calculation. The value of $\mu_{dc}$ was chosen to reproduce the XAS characteristics, shown in the inset to Fig.~\ref{Fig2}(d), and the density of states observed in direct and inverse photoemission~\cite{supplemental}. It is further corroborated by the favorable comparison with ARPES. We thus achieve a reliable model of the electronic structure spanning three orders of magnitude in energy, from the $f-f$ interaction ($U_{ff}=6$\,eV) down to the width of the occupied $f$-states at the Fermi surface ($\sim30$\,meV) and CF splitting (10~meV).

The exact evaluation of the Kramers-Heisenberg term in the AIM comes at the price of losing the momentum-resolution and of discretizing the DFT+{\color{red}DMFT} bath (we use a set of 20 levels~\cite{supplemental}). Subtle aspects of the electronic structure, like an accurate reproduction of the renormalized spin-orbit gap are lost in this necessarily coarse model of local-itinerant hybridization. On the other hand, AIM-RIXS does capture the resonant creation of the excited states. In Fig.~\ref{Fig2}(e), we compare calculated spectra with experimental data above and below the resonance at $E_0$. Both quasiparticle excitations behave Raman-like, i.e., the energy transfer is independent of $h\nu$. Excitations of $f$-quasiparticles that escape into the wide Pd $4d$ bands contribute a fluorescence tail that shifts to higher energies at higher photon energies. This effect is well captured in Fig.~\ref{Fig2}(d), even if the continuum is mapped onto a set of discrete peaks. Thorough discussions of the photon-energy dependence, polarization dependence, and XAS characteristics are provided in the SI~\cite{supplemental}.

\begin{figure}
\includegraphics[width=0.5\columnwidth,trim= 0pt 0pt 0pt 0pt, clip]{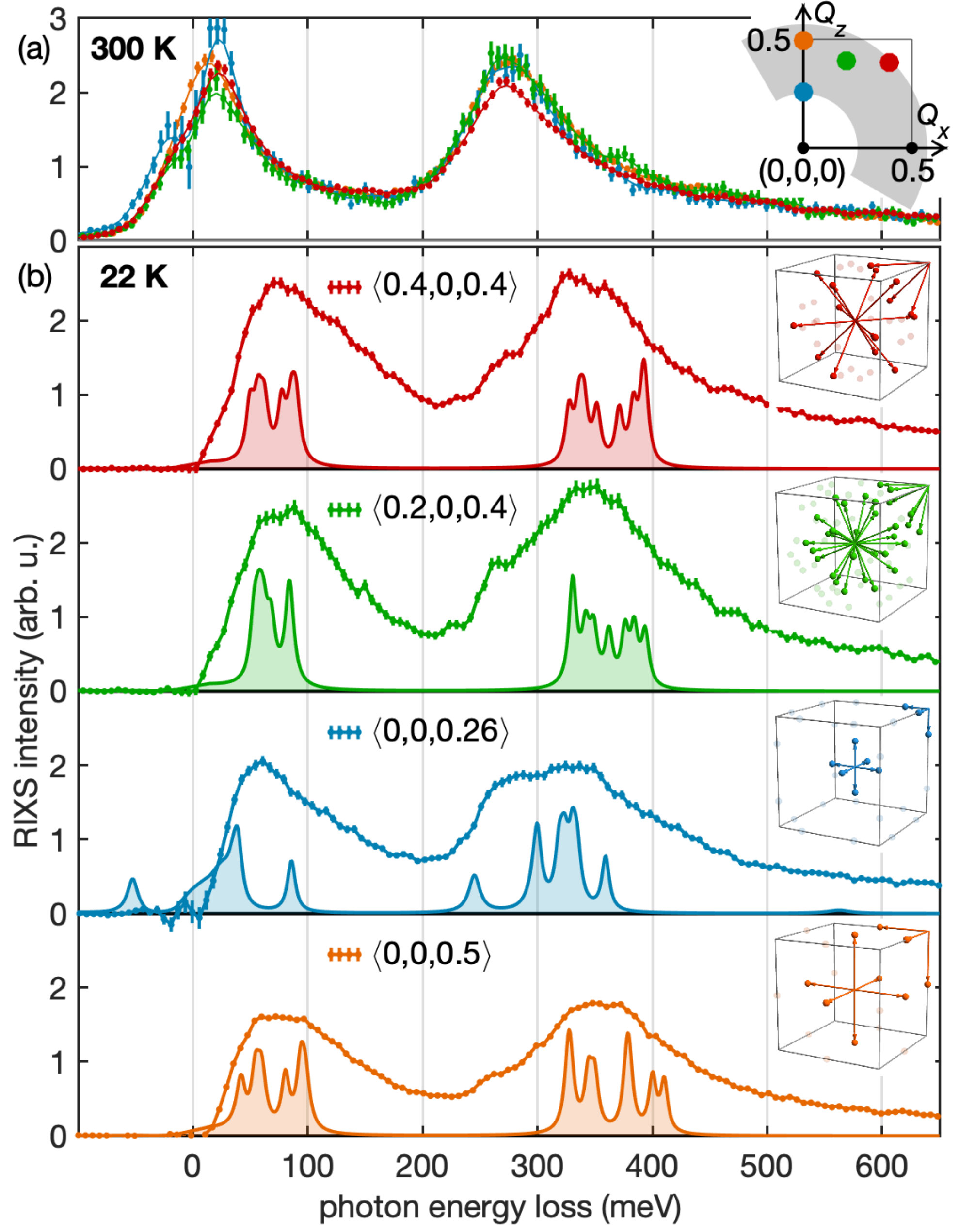}%
\caption{\label{Fig3} \textbf{Fingerprint of a coherent Kondo lattice state in \cpd~as observed by resonant inelastic x-ray scattering (RIXS).} RIXS spectra are shown at four different momentum transfers $Q$, at (a) 300\,K and (b) 22\,K. For better comparison, the quasielastic peak has been subtracted. The positions of $Q$ are indicated in the reciprocal space map inset, where the $Q$ range accessible to $M_5$-edge RIXS is shaded gray. While at $T=$300\,K, well above the coherence temperature $T_\text{coh}\approx130$\,K, the signal is almost invariant with regard to $Q$, for $T=$22\,K a pronounced momentum-dependence emerges. The inset views of the cubic Brillouin zone in (b) illustrate how the $Q$ positions probed with RIXS correspond to momentum transfers exciting heavy quasiparticles from electron pockets at $\Gamma$ and R into unoccupied parts of the electronic structure. For reference, the shaded spectra in each panel show the DFT+DMFT density of states, as shown in Fig.~\ref{Fig1}(c), summed over the relevant positions of quasiparticle momentum-transfers in the Brillouin zone.}
\end{figure}

To probe the momentum-dependence of these $f$-states for $T<T_\mathrm{coh}$, as predicted in Fig.~\ref{Fig1}, we recorded RIXS spectra at four different points in the Brillouin zone, at 300\,K and at 22\,K. As illustrated in Fig.~\ref{Fig3}(a), the response at $T> T_\mathrm{coh}$ is close to isotropic. By contrast, the corresponding spectra in the lattice-coherent state [Fig.~\ref{Fig3}(b)] vary dramatically in both spectral weight and energy. Surprisingly, these observations are most pronounced among the spin-orbit excited quasiparticles, with a dispersion of $\approx40$\,meV between $Q=(0,0,0.26)$ and $(0,0,0.5)$. In the absence of momentum-resolved RIXS calculations, it is instructive to relate these observations to the DFT+DMFT spectra. Because the occupied $f$-quasiparticles are constrained to narrow hot-spots at $\Gamma$ and R, a simple argument can be made: For each of the investigated momentum transfers, symmetry-equivalent excitation paths within the Brillouin zone are illustrated in the insets to Fig.~\ref{Fig3}(b). For instance,  $Q=(0,0,0.5)$ probes transitions ($\Gamma$$\rightarrow$X, R$\rightarrow$M), which should be comparable to $Q=(0.5,0,0.5)$ ($\Gamma$$\rightarrow$M, R$\rightarrow$X). This is consistent with the observation of similar spectra at $Q=(0,0,0.5)$ and $Q=(0.4,0,0.4)$, and significant variations between $Q=(0,0,0.26)$ and $(0,0,0.5)$. For direct comparison, the sum of the 116\,K DFT+DMFT density of states at the respectively relevant positions in the Brillouin zone is drawn as shaded line shapes in Fig.~\ref{Fig3}(b). Despite $E_\mathrm{K}\gg E_\mathrm{CF}$, RIXS is clearly able to resolve the fine structure of the hybridized spin-orbit multiplets.

\begin{figure*}
\includegraphics[width=0.5\columnwidth,trim= 0pt 0pt 0 0pt, clip]{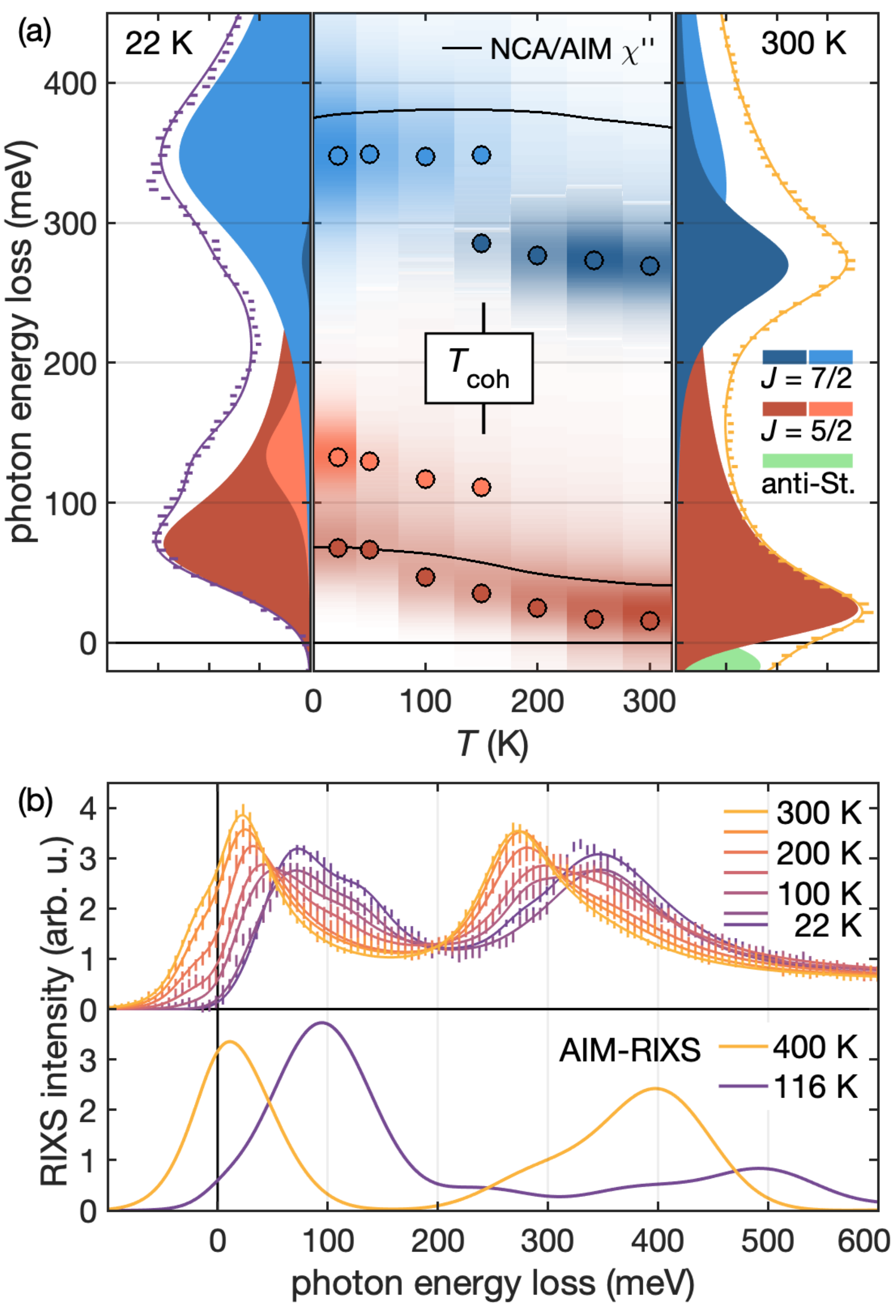}%
\caption{\label{Fig4} \textbf{Onset of coherence in \cpd~as observed by resonant inelastic x-ray scattering (RIXS).}  (a) Detailed thermal variation of RIXS spectra between 22\,K and 300\,K. For clarity, the elastic lines have been subtracted from the data. The emergence of the coherent state at  $T_\mathrm{coh}$ appears as a transfer of intensity between distinct excitations. We phenomenologically model this by Lorentzian fits of the data, as indicated in the left (22\,K) and right (300\,K) panels. Colored markers in the center panel indicate the centroids of these features. As the $J=5/2$ quasiparticles become thermally populated, an anti-Stokes peak (green) appears on the photon-energy-gain side. The black lines show the thermal evolution of the $J=5/2$ and $J=7/2$  peaks in our NCA/AIM calculation of $\chi''$ (see text for details). (b) Comparison of temperature dependent RIXS (top) with the exact AIM-RIXS calculation (bottom). This model accurately captures the parallel shift of $J=5/2$ and $J=7/2$ features. The overestimation of the SO gap is an artifact of the necessary discretization (see text for details).}
\end{figure*}

As shown in Fig.~\ref{Fig4}, the characteristics of the RIXS Kondo excitations change dramatically when thermal fluctuations destroy the lattice-coherence. Fig.~\ref{Fig4}(a) shows spectra at momentum transfer $Q=(0.4,0,0.4)$, at 22\,K and at 300\,K [left and right panels]. Colored lineshapes illustrate phenomenological fits of Lorentzian lineshapes (for better comparison with AIM-RIXS spectra, the elastic line has been subtracted). The center panel shows how, at low temperatures, both excitations continuously shift to higher energies. The black lines indicate the maxima of the momentum averaged dynamic magnetic susceptibility $\chi''(\omega)$ of the Anderson impurity model in the non-crossing approximation (NCA/AIM, see Supplementary~Note~1~\cite{supplemental}). It is thus evident that even phenomenological models reproduce the basic thermal trend of the $J={5/2}$ excitations. However, at intermediate temperatures, both $J={5/2}$ and ${7/2}$ bands develop the momentum-dependent substructure discussed above, which corresponds to the onset of coherence. This microscopic observation of $T_\text{coh}\approx130$\,K is consistent with bulk methods~\cite{Lawrence1985}. Fig.~\ref{Fig4}(b) shows that our AIM-RIXS approach does indeed capture the parallel thermal shift of both SO states by $\approx80$\,meV. The apparent overestimation of the SO gap is due to the limited number of bath levels determined by computational cost, which does not allow to accurately capture the renormalization of $J=7/2$ and $5/2$ due to hybridization with conduction bands.

Our study showcases that RIXS provides a comprehensive picture of the emergent coherent Kondo lattice state, enabling a detailed comparison with material-specific state-of-the-art band structure calculations on all relevant energy scales. Crucially, RIXS reveals the exact position of all $f$-electron levels and how they blend into the underlying metallic state. We have used this capability to clarify the characteristic energy scales of the model Kondo lattice material \cpd. The fine structure and momentum dependence of the observed spectra is consistent with the five CF multiplets  inferred from our DFT+DMFT calculation [Fig.~\ref{Fig1}(a)]. Below the coherence temperature $T_\textrm{coh}\approx130$\,K, which we have determined microscopically, the hybridization with the conduction electrons impacts all $f$ states on the scale of $E_\mathrm{K}\approx55\,$meV. Even though the overall 80\,meV renormalization of the $J={5/2}$ peak as a function of temperature in Fig.~\ref{Fig4}(a) amounts to 30\% of the free-ion spin-orbit coupling (280\,meV), the SO gap remains unchanged in the coherent Kondo lattice state. This implies that the strength of the SO interaction determined by $\mathbf{L}\cdot\mathbf{S}$ is not impacted by the substantial renormalization of the electronic state that is evidenced by overall shifts of the $f$-levels (cf. Fig.~\ref{Fig4}) and their variation throughout the Brillouin zone (cf. Fig.~\ref{Fig1}), both of which are on the order of $E_\textrm{K}$. Interestingly, this observation is contrary to the interpretation by Murani~\textit{et al.}~\cite{Murani2002b}, who expected that $f$ band formation should lead to at least a partial quenching of orbital momentum and a reduction of $\mathbf{L}\cdot\mathbf{S}$.

In conclusion, our study reveals in unprecedented detail how the full manifold of localized $f$-electron states becomes hybridized with the underlying band structure in the archetypal Kondo metal \cpd. Such insight is crucial for the development of fully microscopic and material-specific models of quantum-coherency and to understand how it drives the emergence of novel order parameters. The remarkable discovery that $f$-electron band formation below the coherence temperature does not quench the spin-orbit interaction showcases how ---via the mechanism of hybridization--- even excited $f$-electron states can endow a material with novel properties that cannot be derived simply from the symmetry of the underlying crystal lattice. This is key to a growing number of topical phenomena such as $p$-wave superconductivity, electronic-nematic order, and topologically non-trivial phases. Given that the present insights were not limited by the experimental energy resolution, similar studies are already feasible in materials with lower Kondo energy scales or in systems where Kondo dynamics can be tuned by external parameters. They can also be performed on microscopic single crystals not accessible to INS and will benefit from continuous improvements in brilliance and energy resolution~\cite{Dvorak2016}. 

\clearpage

\begin{center}
{\bf Methods}
\end{center}
\small
\begin{center}
{Samples}
\end{center}
An ingot of cubic \cpd~(AuCu$_3$ structure, space group $Pm\bar{3}m$, lattice parameter $a=4.13$\,\AA) was grown from a self-contained melt using a modified Czochralski method in a tri-arc furnace. Single-crystalline grains were identified and oriented by backscattering Laue x-ray diffraction. Cuboid samples with faces parallel to [100] planes were then cut from this grain. Incisions were made around the circumference of these samples to favor fracturing parallel to these planes.

\begin{center}
{Photoemission spectroscopy}
\end{center}
Vacuum ultraviolet angle-resolved photoemission spectroscopy was performed at a BL5-2, SSRL (Stanford), beamline 4.0.3, ALS (Berkeley), and one-cube, BESSY-II (Berlin). The samples were fractured under ultra-high vacuum conditions of $<5\times10^{-11}$\,Torr. The best contrast from the Ce $4f$ states was obtained above the N ($4d$$\rightarrow$$4f$) absorption edge, at photon energies of 130$\sim$150\,eV. The electrons were analyzed using Scienta R8000 (4.0.3) and Scienta DA30L (BL5-2, one-cube) spectrometers. 

\begin{center}
{X-ray spectroscopy}
\end{center}
Resonant inelastic x-ray scattering experiments were carried out at endstation ID32-ERIXS, ESRF (Grenoble), with a combined energy resolution of 34\,meV at the Ce $M_5$ resonance ($h\nu\approx882$\,eV)~\cite{Brookes2018}. The samples were fractured in-situ under ultra-high vacuum conditions. Data was collected in cycles of 5\,min. for 7--8\,h. per spectrum and processed using the RIXS Toolbox suite~\cite{Kummer2017}. Data shown in the main text is measured with incident polarization in the scattering plane, without polarisation analysis of the scattered beam. The polarization analysis of the scattered beam~\cite{Braicovich2014} is demonstrated in Supplementary~Note~6~\cite{supplemental}. Ce $M$ edge x-ray absorption spectra were obtained during the RIXS experiment  (total electron yield mode) and were confirmed by additional measurements in fluorescence yield mode measured at BL29/BOREAS, ALBA (Barcelona), see~\cite{supplemental}.

\begin{center}
{Inelastic neutron scattering}
\end{center}
Neutron time-of-flight neutron spectra of CePd$_3$ were obtained on ARCS, SNS (Oak Ridge) at an incident energy of $E_i= 400$\,meV. The  same  20\,g  single  crystal and  experimental procedures were used as in Ref.~\cite{Goremychkin2018}.

\clearpage
\begin{center}
{Computational models}
\end{center}
The temperature evolution of the $^2F_{5/2}$ and $^2F_{7/2}$ excitations (Fig.~\ref{Fig4}(a), black solid lines) was inferred by the momentum-averaged dynamic susceptibility $\chi''(\omega)$ in the Anderson impurity model (AIM) computed with the non-crossing approximation~\cite{Bickers1987b}, as described in our earlier work~\cite{Lawrence2001} and discussed in~\cite{supplemental}. The NCA/AIM description assumes a single $f$-level (with large orbital degeneracy), a constant hybridization with a Gaussian shape (fixed width) of the conduction band.

The DFT+DMFT electronic structures of CePd$_3$ shown in Fig.~\ref{Fig1} were obtained using the Wien2K package~\cite{Blaha2018,Blaha2020} and the strong-coupling continuous-time quantum Monte Carlo impurity solver~\cite{Werner2006,Boehnke2011,Hafermann2012,Hariki2015}, details can be found in~\cite{supplemental}. The cerium site energy was shifted by the double-counting term $\mu_{dc}$, which corrects for the $f$-$f$ interaction inherent in the LDA results. The $\mu_{dc}$ value was chosen to reproduce the experimental XAS [cf.~Fig.~\ref{Fig2}(d)], as well as the electronic spectral function $A(k,\omega)$ observed in direct/inverse photoemission and ARPES data of CePd$_3$, as illustrated in Figs.~\ref{Fig1}(f), \ref{Fig2}(b) and Refs.~\cite{supplemental,Malterre1992,Souma2001}.
We employed a diagonalization solver with a configuration-interaction scheme for evaluating the RIXS intensities using the Kramers-Heisenberg term from the AIM with the DFT+DMFT hybridization densities (Figs.~\ref{Fig2}(d) and \ref{Fig2}(e)). The AIM-RIXS calculations of the density of states, XAS and RIXS spectra were performed in analogy to our earlier work~\cite{Hariki2020} and are discussed in detail in~\cite{supplemental}.



\begin{acknowledgements}
We are grateful for provision of experimental time at beamline ID32 of the European Synchrotron Radiation Facility (ESRF), Grenoble, France, at beamline BL5-2 of the Stanford Synchrotron Radiation Lightsource (SSRL) of the SLAC National Accelerator Laboratory, operated by the U.S. Department of Energy (DOE), Office of Science, Office of Basic Energy Sciences (BES) under Contract No. DE-AC02-76SF00515, at beamline 4.0.3. of the Advanced Light Source, a U.S. DOE Office of Science User Facility under contract no. DE-AC02-05CH11231, at beamline UE112\_PGM-2b-1\textasciicircum 3 of the BESSY-II electron storage ring operated by the Helmholtz-Zentrum Berlin f\"{u}r Materialien und Energie (HZB), at beamline BL29 of the ALBA Synchrotron, Spain, and at the ARCS spectrometer at the Spallation Neutron Source, a DOE Office of Science User Facility operated by the Oak Ridge National Laboratory. Work at Los Alamos National Laboratory was performed under the auspices of the US Department of Energy, Office of Basic Energy Sciences, Division of Materials Sciences and Engineering. M.C.R. is grateful for support through the LANL Director's Fund and the Alexander von Humboldt Foundation. J.-X.Z. was supported by the Center for Integrated Nanotechnologies, a DOE BES user facility. A.H., K.-H.A. and J.K. were supported by the European Research Council (ERC) under the European Union’s Horizon 2020 research and innovation programme (grant agreement No. 646807-EXMAG). A.H. was supported by JSPS KAKENHI Grant Number 21K13884. K.-H.A. was supported by the Czech Science Foundation Project Grant Number 19-06433S. A.D.C. was partially supported by the U.S. Department of Energy, Office of Science, Basic Energy Sciences, Materials Sciences and Engineering Division. A.A. acknowledges the financial support of the Deutsche Forschungsgemeinschaft DFG under project SE1441/5-2. Work at TU Dresden was supported by the Deutsche Forschungsgemeinschaft through the CRC 1143 and the Würzburg-Dresden Cluster of Excellence ct.qmat (EXC2147, Project ID 390858490). The work by F.H. and M.J. was supported through a Hans Fischer fellowship of the Technische Universität München—Institute for Advanced Study, funded by the German Excellence Initiative and the European Union Seventh Framework Programme under Grant agreement No. 291763. Special thanks are due to Daniel Mazzone, Johan Chang, Pascoal Pagliuso, Jason Hancock and Peter Riseborough for insightful discussions.
\end{acknowledgements}

\clearpage
\begin{center}
{\bf Author contributions}
\end{center}
M.C.R., M.J., F.R., and J.M.L. designed the research. M.C.R., E.B., D.D.B, and K.J.M. synthesized the samples. M.C.R. planned and conducted the RIXS and ARPES measurements, with assistance from M.J., A.A., and F.H.; K.K. and J.D.D. performed additional RIXS, XAS and ARPES measurements. A.D.C. and J.M.L. planned and performed the neutron measurement. Instrument scientists K.K., J.D.D., D.-H.L., M.H., E.R., and M.V. enabled and assisted in experiments. A.H., J.K., K.-H.A., J.-X.Z., and C.H.B. performed calculations and advised on the theoretical interpretation of the data. M.C.R. analyzed the data. M.C.R and M.J. wrote the manuscript with feedback from all authors.

\clearpage

\clearpage


\section*{Supplementary material}

# Supplementary Information

## Kondo quasiparticle dynamics observed by resonant inelastic x-ray scattering

M. C. Rahn, K. Kummer, A. Hariki, K.-H. Ahn, J. Kuneš, A. Amorese, J. D. Denlinger, and D.-H. Lu, M. Hashimoto, E. Rienks, M. Valvidares, F. Haslbeck, D. D. Byler, K. J. McClellan, E. D. Bauer, J.-X. Zhu, C. H. Booth, A. D. Christianson, J. M. Lawrence, F. Ronning, and M. Janoschek

**Supplementary Notes**

## Supplementary Note 1: NCA/AIM calculations of $\chi''$

Calculations of the dynamic magnetic susceptibility $\chi''$ in the Anderson impurity model were performed using the large degeneracy or non-crossing approximation [1], as described in our earlier work [2]. This "toy model" of the hybridization phenomenon assumes a Gaussian conduction band of width $W$ centered at the Fermi surface $E_F$, interacting with $4f^1$ states at the binding energy $E_f$, with a hybridization constant $V$, and spin-orbit–coupling $\Delta_{SO}$. The calculations yield both static and dynamic quantities, such as the $f$-occupation number $n_f$, magnetic susceptibility $\chi$, and powder-averaged $\chi''$, and has proven universal in its applicability to $CePd_3$ [3] and other intermediate valence materials [4].

The input values for the background bandwidth $W$, the hybridization constant $V$, and the $f$-level energy $E_f$ are not obtained from first-principles calculations, nor do they necessarily reflect values taken from experiment. Rather, the calculation remains valid for the given experimental measurements as long as the universality of the Kondo physics holds, where the given quantities are universal functions of $T/T_K$ or $E/(k_B T_K)$. Since $T_K$ depends exponentially on the ratio $(W E_f/V^2)$, the given parameters will reproduce the data as long as this ratio is approximately correct. To illustrate these characteristics, in Supplementary Fig. 1, we show a comparison of calculations for two sets of parameters.

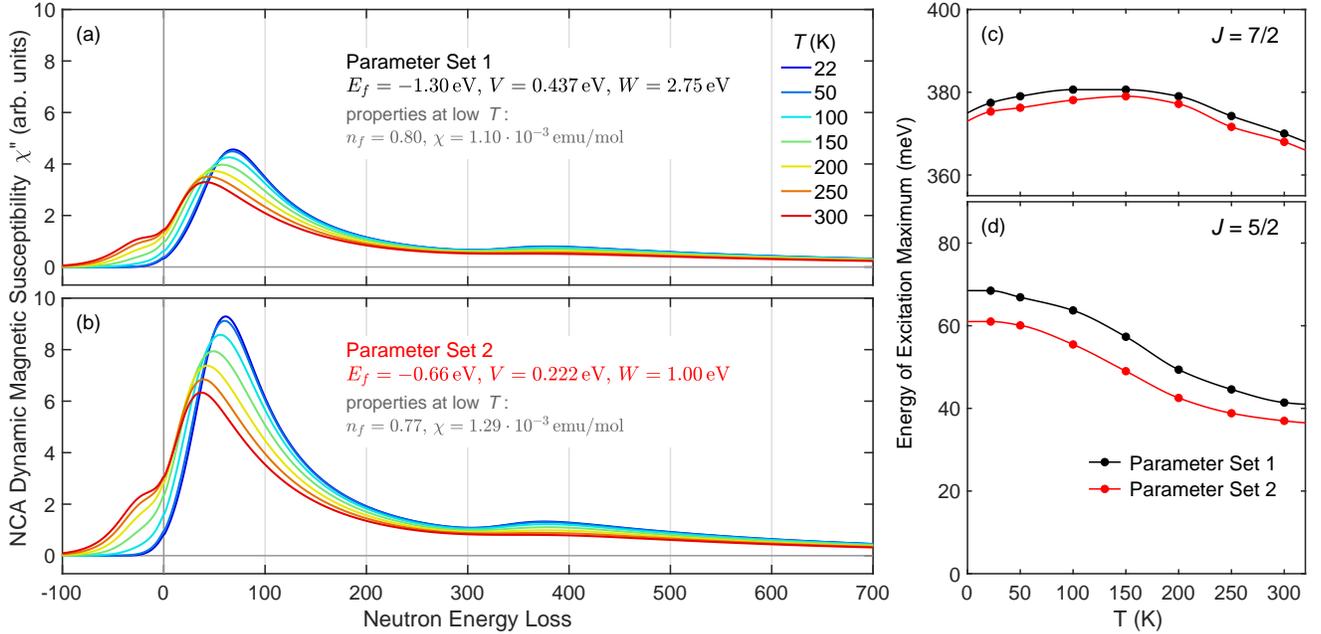

FIG. 1. (a,b) NCA/AIM calculations of the momentum-averaged neutron magnetic scattering function $S(\omega) = [1 + n(\omega)]\frac{1}{\pi}\chi''(\omega)$, for a range of temperatures. The corresponding sets of model parameters $E_f$, $V$ and $W$ are indicated in each panel, along with the resulting values of $n_f$ and $\chi$ at 5 K. (c,d) Position of the corresponding maxima of $J = 5/2$ and $J = 7/2$ excitations. Lines are intended as a guide to the eye.

## Supplementary Note 2: Comparison of RIXS and DFT+DMFT calculations of $\chi''$

Goremychkin *et al.* have calculated the dynamic magnetic susceptibility of $CePd_3$ as the Lindhard susceptibility of a DFT+DMFT simulation, taking into account two-particle vertex corrections $\Gamma^{loc}_{irr}$ [5]. In Supplementary Fig. 2, we present a comparison of these results with RIXS spectra obtained at similar momentum transfers and temperatures (cf. colored lines/data adopted from the Supplementary Materials of Ref. [5]). For reference, we also show the results of an earlier DFT+DMFT calculation by Sakai [6] (dashed line), as well as our AIM/NCA calculation, as discussed above (dot-dashed line).

The direct comparison between RIXS spectra and three computational models of $\chi''$ reveals some interesting similarities and differences. As in our simpler AIM/NCA model, $\chi''$ obtained by DFT+DMFT is dominated by excitations within the $J = 5/2$ ground state manifold and features only a weak broad signal at the spin-orbit energy. However, the DFT+DMFT spectra by Goremychkin *et al.* show little or no shift of the $J = 5/2$ excitation at low temperatures. Instead, an intermediate excitation emerges at around 150–200 meV. We have probed for this excitation using INS with an incident neutron energy of 400 meV [see purple markers in Supplementary Fig. 3(b)] and confirmed its absence. Instead, the $J = 5/2 \rightarrow 5/2$ neutron excitations are in excellent agreement with RIXS. Even though the origin of the 150–200 meV feature in DFT+DMFT is unclear, it shares the dispersive trend of the $J = 5/2 \rightarrow 7/2$ excitation observed in RIXS.

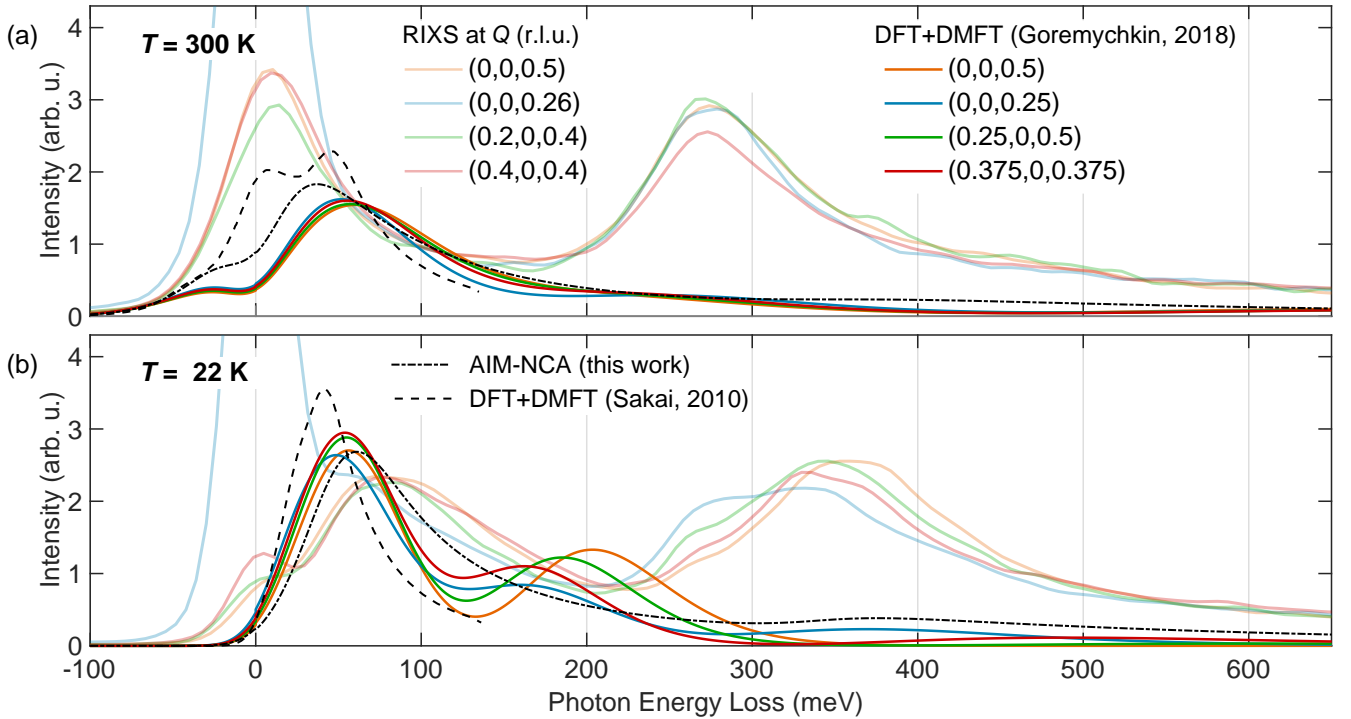

FIG. 2. Comparison of RIXS spectra with DFT+DMFT calculations of the dynamic magnetic susceptibility $\chi''(Q,\omega)$ as reported by Goremychkin *et al.* [5], with a DFT+DMFT calculation of $\chi''(\omega)$ reported earlier by Sakai [6], and with the results of the NCA/AIM "toy model" discussed above. Lines drawn in corresponding colors indicate data measured/calculated at similar momentum transfer. Each set of calculations is arbitrarily scaled to the RIXS data, however with the same scale factor for high and low temperature calculations. Dot-dashed lines reproduce our NCA/AIM model illustrated in Supplementary Fig. 1 (Parameter Set 1).

## Supplementary Note 3: Comparison of RIXS and inelastic neutron scattering

The fact that the RIXS excitations of $CePd_3$ appear in the crossed polarization channel suggests that the dynamic magnetic susceptibility $\chi''(Q,\omega)$, as measured by INS, may provide a useful comparison. We measured the polycrystalline-averaged dynamical correlation function $S(\omega) = [1+n(\omega)]\frac{1}{\pi}\chi''(\omega)$ [ $n(\omega)$: Bose occupation factor] on at ARCS (ORNL), which allows access to the 50–250 meV range. The spectrum is shown in Supplementary Fig. 3(b) along with data reproduced from Fanelli *et al.* [3] (measurement at 7 K), Murani *et al.* [7] (10 K), and Murani *et al.* [8] (12 K). For reference, the RIXS spectra of Fig. 3 of the manuscript are reproduced in Panel (a), and the ranges of these RIXS intensities are also indicated in Panel (b) by gray shaded margins. Where available, the spectra are presented in absolute units.

Remarkably, RIXS and INS are quantitatively consistent up to ∼ 200 meV, i.e., both in the energy position, lineshape, and in the relative intensity between low and high-temperature datasets [3]. On the other hand, RIXS and INS differ markedly at energy transfers above 200 meV. Due to the weakness of dipole transition matrix elements between the $J = 5/2$ and $J = 7/2$–like states, excitations across the spin-orbit gap are hardly measurable in INS. For reference, the red markers in Supplementary Fig. 3(b) show results of the epicadmium (very high incident energy) INS study by Murani *et al.*, Ref [8]. After subtracting a $J = 5/2$ Lorentzian tail, the authors identified a weak spin-orbit feature, shown here enlarged by a factor 180.

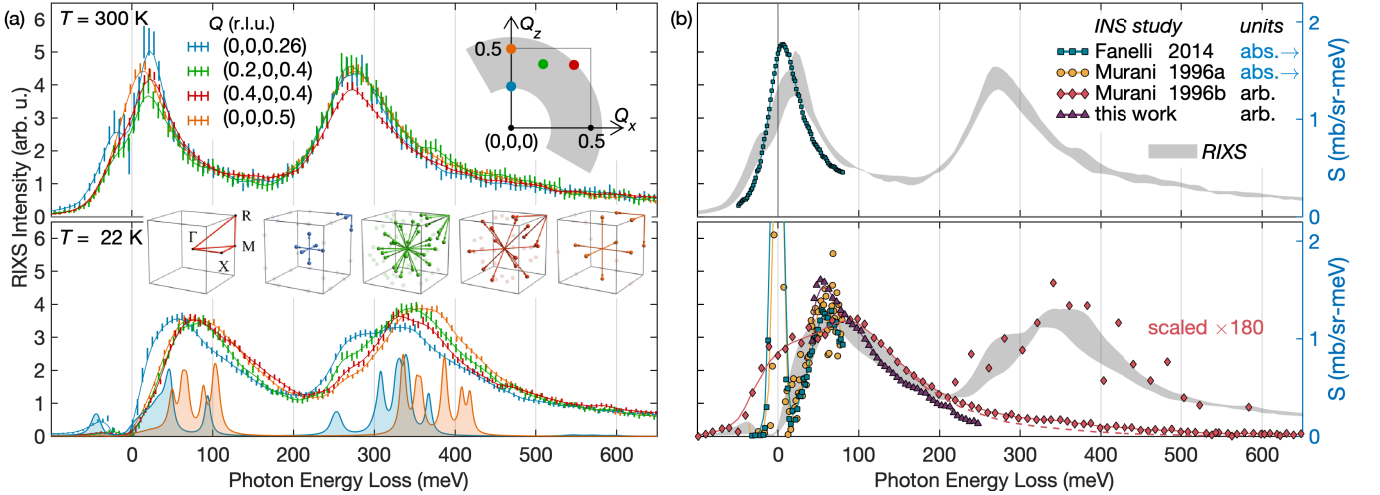

FIG. 3. Emergent RIXS momentum-dependence and comparison with neutron scattering. (a) RIXS spectra at four different momentum transfers $Q$, at 300 K (top) and 22 K (bottom). The positions of $Q$ are indicated in the reciprocal space map inset, where the $Q$ range accessible to $M_5$-edge RIXS is shaded gray. The lower inset illustrates the corresponding momentum transfers to quasiparticles in the Fermi surface pockets at $\Gamma$ and R. For the extremal cases, $Q = (0, 0, 0.26)$ and $(0, 0, 0.5)$, the DFT+DMFT density of states summed over the relevant positions in the Brillouin zone are drawn as shaded peaks in corresponding colors. (b) Quantitative comparison of the RIXS spectra presented in Panel (a), here drawn as shaded margins, with polycrystalline-averaged neutron spectra of $CePd_3$ reported by Fanelli *et al.*, Ref. [3] and Murani *et al.*, Ref. [7, 8]. The neutron data marked *this work* has not been previously published. Datasets in which the neutron scattering function $S$ is available in absolute units (mb/sr-meV) are marked [abs.] in the legend (cf. right axis).

**Supplementary Note 4: DFT+DMFT AIM-RIXS calculation of the Kramers-Heisenberg term**

**4.1 Overview of the Method**

For our computation of Ce $M_5$-edge RIXS spectra using the DFT+DMFT scheme, we proceed in several steps [9–15]. First, a standard DFT+DMFT calculation is performed for the experimental crystal structure of CePd$_3$, as follows. The DFT bands within the generalized gradient approximation (GGA) [16] for the exchange-correlation potential are obtained using the WIEN2K package [17], which implements the augmented plane wave and the local-orbital (APW+lo) method. Muffin-tin radii (RMT) of 2.5 $r_{\text{Bohr}}$ are used for Ce and Pd atoms. The maximum modulus for the reciprocal vectors $K_{\max}$ was chosen such that $R_{\text{MT}} \times K_{\max} = 7.0$. The Brillouin zone was sampled in a $20 \times 20 \times 20$ $k$-mesh. The spin-orbit coupling (SOC) is also taken into account in the GGA calculations.

Second, the DFT bands are projected onto a tight-binding model spanning Ce $4f$ and $5d$ bands, and Pd $4d$, $5s$, and $5p$ bands using the wien2wannier [18] and wannier90 [19] codes. This tight-binding model is augmented by the local electron-electron interaction $U$ within the Ce $4f$ shell. Following previous DFT+DMFT and spectroscopy studies for Ce compounds [5, 20], we employed $U = 6$ eV for the Hubbard parameter. The bare energy of the Ce 4f states is obtained from the GGA value by subtracting the double-counting correction $\mu_{\text{dc}}$, which accounts for the Ce $4f$–$4f$ interaction already present in the GGA description. In the absence of a unique definition of $\mu_{\text{dc}}$, we treat $\mu_{\text{dc}}$ as a parameter, adjusted by comparison to experimental valence and inverse photoemission spectra [21, 22] (see below / Supplementary Fig. 5).

Finally, to compute Ce $M_5$-edge RIXS and XAS spectra using the Kramers-Heisenberg formula [23] and Fermi's golden rule, we adopt the configuration-interaction (CI) solver [13, 15, 24]. This AIM solver uses the hybridization density $V^2(\varepsilon)$ obtained from DMFT. The AIM is augmented by the Ce $3d$ core states and the $3d$–$4f$ interaction present in the intermediate state of RIXS (corresponding to the final state of XAS). The core-hole potential $U_{fc}$ is set to 8.8 eV, and the higher Slater integrals obtained from atomic Hartree-Fock calculation [25] are scaled down to 80 % of their actual values to simulate the effect of intra-atomic configuration interaction from higher basis configurations neglected in the atomic calculation [25–27].

As the CI solver works directly in real frequencies, it allows to resolve fine spectral features in the spectra. However, it needs to approximate the continuum hybridization densities $V^2(\varepsilon)$ by a finite set of discrete bath levels. This CI method allows us to include more bath levels than in a standard exact diagonalization algorithm (see recent studies for $3d$ transition metal oxides [15, 24]). Nevertheless, for the case of a Ce $4f$ impurity with a large number of internal (local) degrees of freedom, the necessarily finite number of bath levels still poses a severe limitation.

## 4.2 Projection of bands onto a local basis

The Wannier functions $w(\mathbf{r})$ representing the Ce $4f$ states in the tight-binding model are summarized in Supplementary Fig. 4. In projecting the Ce $4f$ bands, we adopt the $\Gamma_{6,7,8}$ basis, i.e. these Wannier functions are constructed directly as spinors (with different up and down components). The spin components of the Ce $4f$ Wannier functions are displayed in Supplementary Fig. 4. The local Hamiltonian at the Ce site can be found in the table shown in this figure, where small off-diagonal matrix elements are allowed between the $\Gamma_7 - \Gamma_7'$ and $\Gamma_8 - \Gamma_8'$ states.

Since our Ce $4f$ Wannier functions are an irreducible representation of the cubic symmetry, the off-diagonal hybridization densities $V^2(i\omega_n)$ [discussed in Supplementary Note 4.4] are allowed only between the states belonging to the same (two $\Gamma_7$ or two $\Gamma_8$) representations. Given that the two $\Gamma_7$ (or $\Gamma_8$) states are well split by the spin-orbit coupling, the resulting off-diagonal hybridization densities are small [cf. Supplementary Fig. 6(c)].

To model the electron-electron interaction in the AIM calculation, these small off-diagonal hybridization densities are neglected, and only the isotropic Hubbard term $U$ is taken into account. Using this computational setup, the CT-QMC simulation can access low temperatures, free from the sign problem.

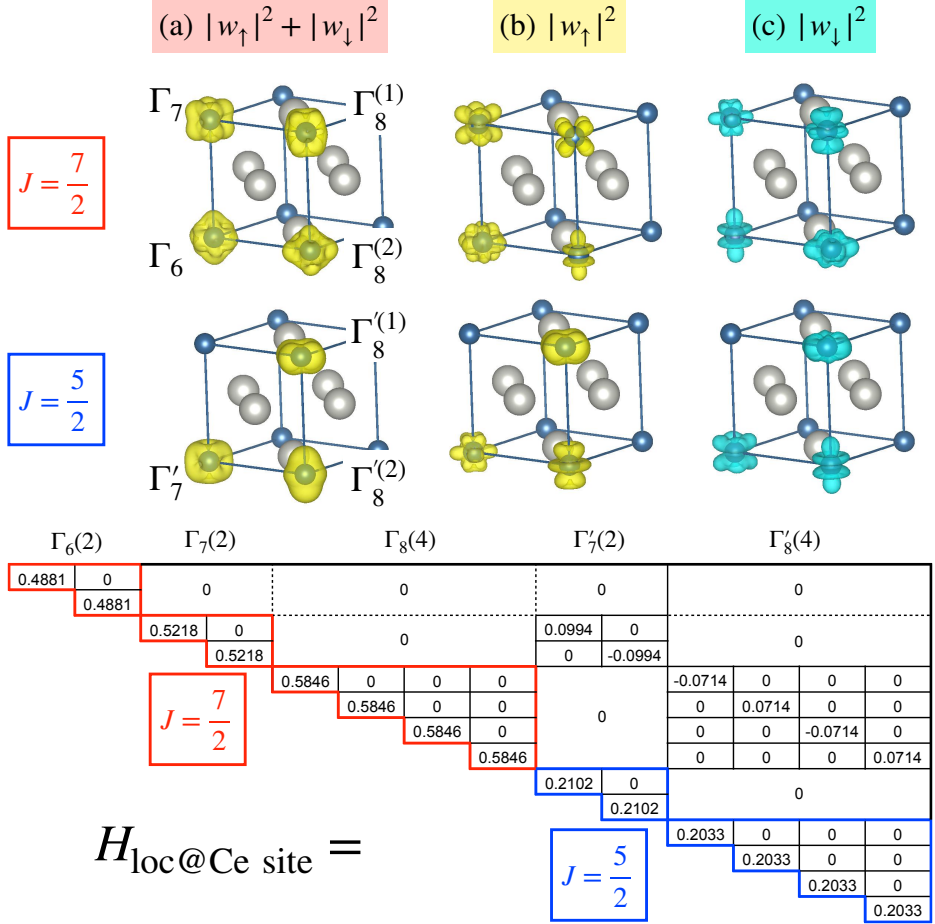

FIG. 4. The density of (a) the Ce 4f Wannier functions ($|w_\uparrow|^2 + |w_\downarrow|^2$), (b) the up-spin components ($|w_\uparrow|^2$), and (c) the down-spin components ($|w_\downarrow|^2$) of the wannier functions. The local Hamiltonian $H_{\text{loc}}$ (the upper triangle part) at the Ce site is shown in the table.

### 4.3 The double counting parameter $\mu_{\text{dc}}$

As stated above, the bare energy of the Ce $4f$ states is obtained from the GGA value by subtracting the double-counting correction $\mu_{\text{dc}}$, which accounts for the Ce $4f$–$4f$ interaction that is already present in the GGA description. In practice, $\mu_{\text{dc}}$ renormalizes the energy splitting between the Ce $4f$ states and Pd $4d$ (and other) bands, which is illustrated in Supplementary Fig. 5. As a consequence, the position of the $4f^2$ peak (at around 5 eV in the experimental data) is sensitive to $\mu_{\text{dc}}$, despite a small change in the Ce $4f$ occupation ($N_{4f}$ in the figure). This allows us to fix $\mu_{\text{dc}} \approx -2.25$ eV.

Aside from these valence spectra (Supplementary Fig. 5), the obtained Ce $4f$ self-energies $\Sigma(\varepsilon)$ are also highly sensitive to $\mu_{\text{dc}}$, as discussed below in the context of Supplementary Fig. 6.

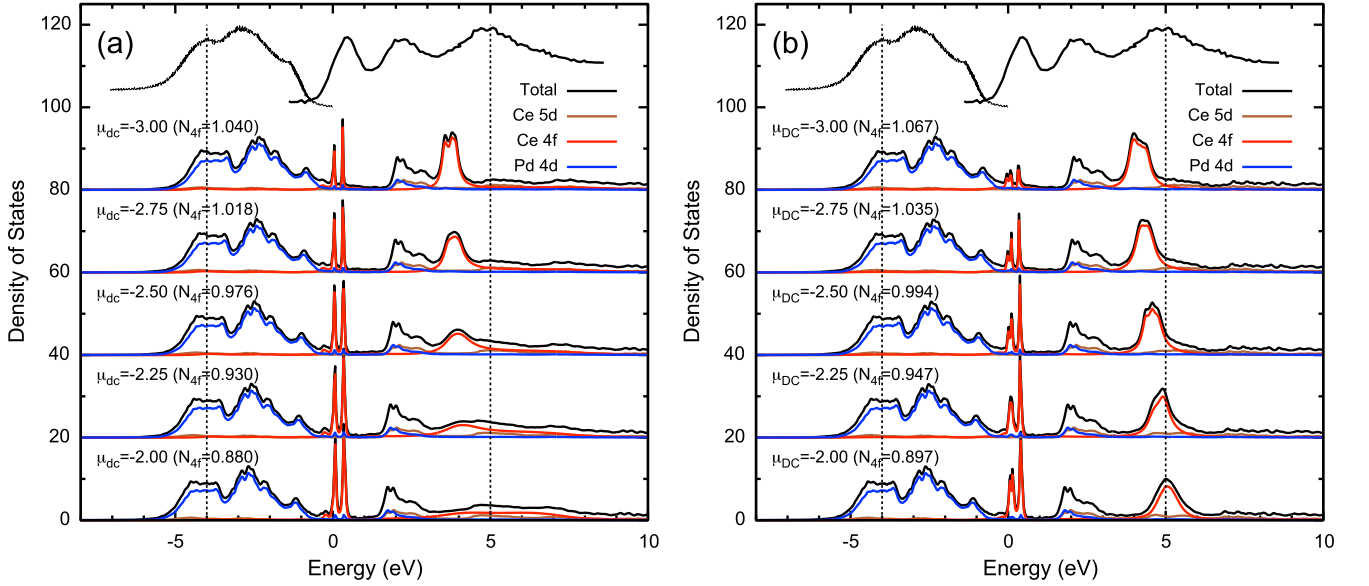

FIG. 5. The double-counting ($\mu_{\text{dc}}$) dependence of the valence spectra at (a) low (116 K) and (b) high (400 K) temperatures calculated by the DFT+DMFT method. The experimental photoemission and inverse photoemission spectra are taken from Refs. [21, 22].

### 4.4 Self energies $\Sigma(\varepsilon)$

In DMFT, the electron dynamics on the Ce sites is described by the Anderson impurity model (AIM) [9, 10]. The exchange of the electrons between a Ce site and the rest of the crystal (represented by non-interacting auxiliary bath states) is described by the hybridization density $V^2(i\omega_n)$ that is determined self-consistently within DMFT. We use a strong-coupling continuous-time quantum Monte Carlo (CT-QMC) impurity solver [28–31] to compute the self-energies $\Sigma(i\omega_n)$ from the AIM.

Aside from the valence spectra (Supplementary Fig. 5), the Ce $4f$ self-energies $\Sigma(\varepsilon)$ also show a strong $\mu_{dc}$ dependence, see Supplementary Fig. 6. This relates with the physics of a highly asymmetric Anderson impurity model, where the $4f^2$ state (upper Hubbard band) lies at around 5 eV, while the $4f^0$ state (lower Hubbard band) is close to the chemical potential. In terms of ionic configurations, the $f^0$ state is rather close to the $f^1$ ground state, and thus even a small shift of the bare $4f$ levels (by the double-counting correction $\mu_{dc}$) yields a large change of $E(f^1) - E(f^0)$. This also affects the Kondo temperature. Consequently, the thermal evolution of both the low-energy Ce $4f$ spectra (Supplementary Fig. 5) and the self energy (Supplementary Fig. 6) depend on $\mu_{dc}$.

Given this $\mu_{dc}$-dependence, we do not implement the charge self-consistency (updating the DFT charge) in our DFT+DMFT calculation. Instead we adjust $\mu_{dc}$ as a key parameter to achieve consistency with experimental results. This includes previously reported (direct and inverse) photoemission spectra (Supplementary Fig. 5), as well as the spectroscopic results (ARPES, $M$-edge XAS, RIXS) reported in the manuscript.

After reaching DMFT self-consistency, we analytically continue $\Sigma(i\omega_n)$ with the maximum entropy method [32, 33] to $\Sigma(\varepsilon)$ in real frequencies $\varepsilon$. $\Sigma(\varepsilon)$ is then used to calculate the one-particle spectral densities $A(\varepsilon)$ and hybridization densities $V^2(\varepsilon)$ at real frequencies, as shown in Supplementary Fig. 6(c).

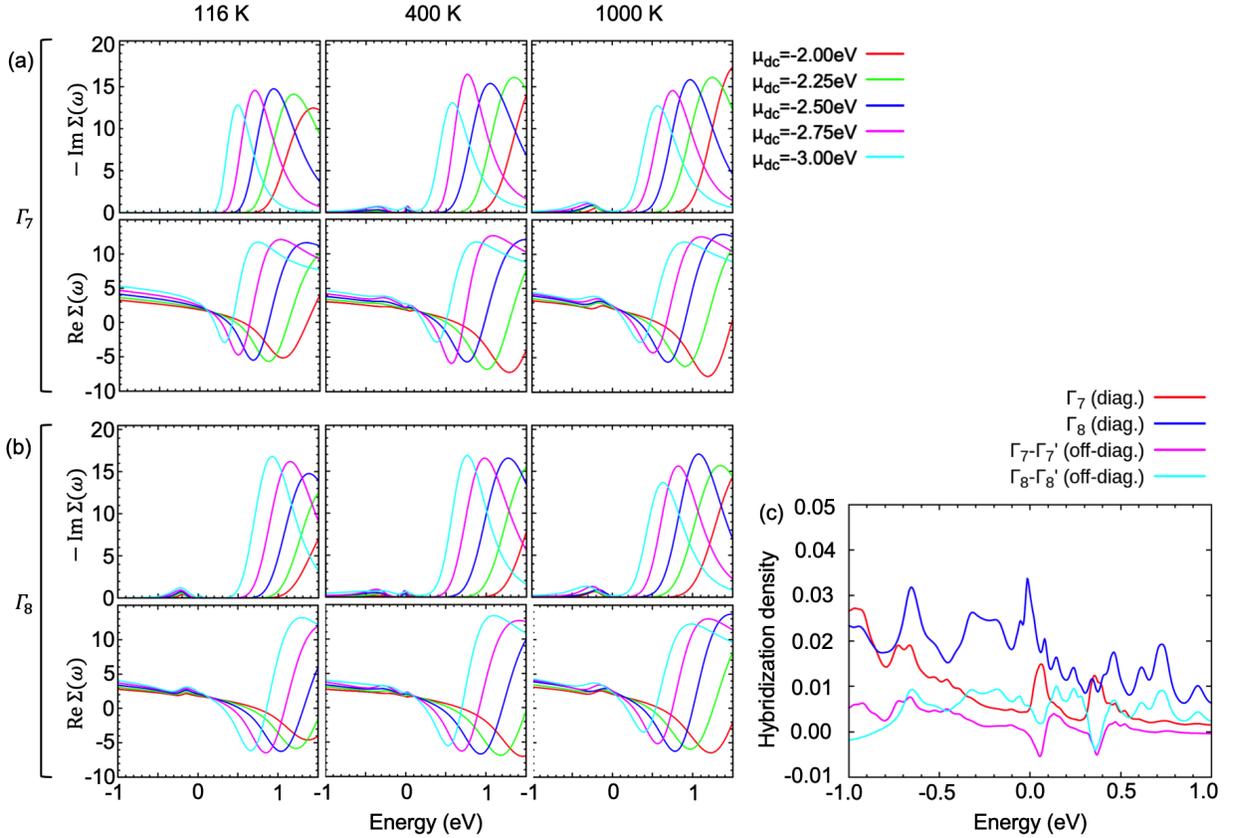

FIG. 6. Real and imaginary parts of the self energy in CePd$_3$, calculated for different double counting parameters $\mu_{dc}$, at 116 K, 400 K, and 1000 K, (a) for $\Gamma_7$ orbitals and (b) for $\Gamma_8$ orbitals. (c) Hybridization density for $\Gamma_7$ and $\Gamma_8$, as well as for off-diagonal terms.

8

## 4.5 Discretization of hybridization densities

As noted above, the AIM solver works directly in real frequencies. This allows to resolve fine spectral features in the spectra, but requires to approximate the continuum hybridization densities $V^2(\varepsilon)$ by discrete levels. In the present study, we modeled $V^2(\varepsilon)$ by 20 bath levels in the discretization scheme $[-5:1:-1,-0.08:0.02:0,0.25:0.25:2.5]$ eV, as illustrated below.

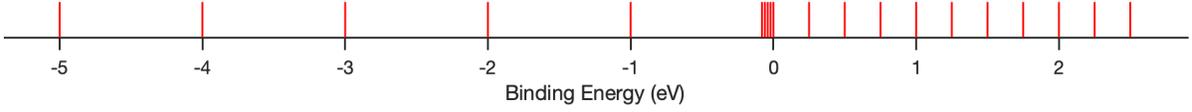

The CI implementation of the present discrete bath gives an accurate $4f$ occupation $N_{4f}$ (the deviation from the numerically-exact one estimated by CT-QMC is less than 5%) for the used $\mu_{\rm dc}$ value ($-2.25$ eV) and simultaneously reproduces the energy of the $J=5/2 \to 5/2$ transitions ($\sim 70$ meV, see the main text). In testing various discretization schemes of $V^2(\varepsilon)$, we found that only those with an accurate $N_{4f}$ value also give a good match in the position of the $J=5/2 \to 5/2$ peak. At high temperatures, low-energy RIXS features are basically atomic excitations involving the $J=7/2$ and $J=5/2$ manifolds, i.e., the given Ce site and the rest of the crystal are effectively decoupled from each other. Thus the details in the bath discretization do not affect the low-energy RIXS features.

**Supplementary Note 5: RIXS incident energy dependence and Pd $4d$ bands**

Supplementary Fig. 7 shows the variation of RIXS spectra of CePd$_3$ in the vicinity of the Ce M$_5$ absorption edge. The measurements were obtained at room temperature, by exciting the sample with x-rays linearly polarized in the scattering plane, and with polarization-resolution of the scattered beam ($\pi\pi'+\pi\sigma'$ polarisation channels). The data allows distinguishing a broad fluorescence-like signal with energy-transfers proportional to the excitation energy ($E \propto \hbar\omega_{\text{in}}$) from the Raman-like ($E = $const.) interband excitations. The right panel shows a detailed view of low energy transfers.

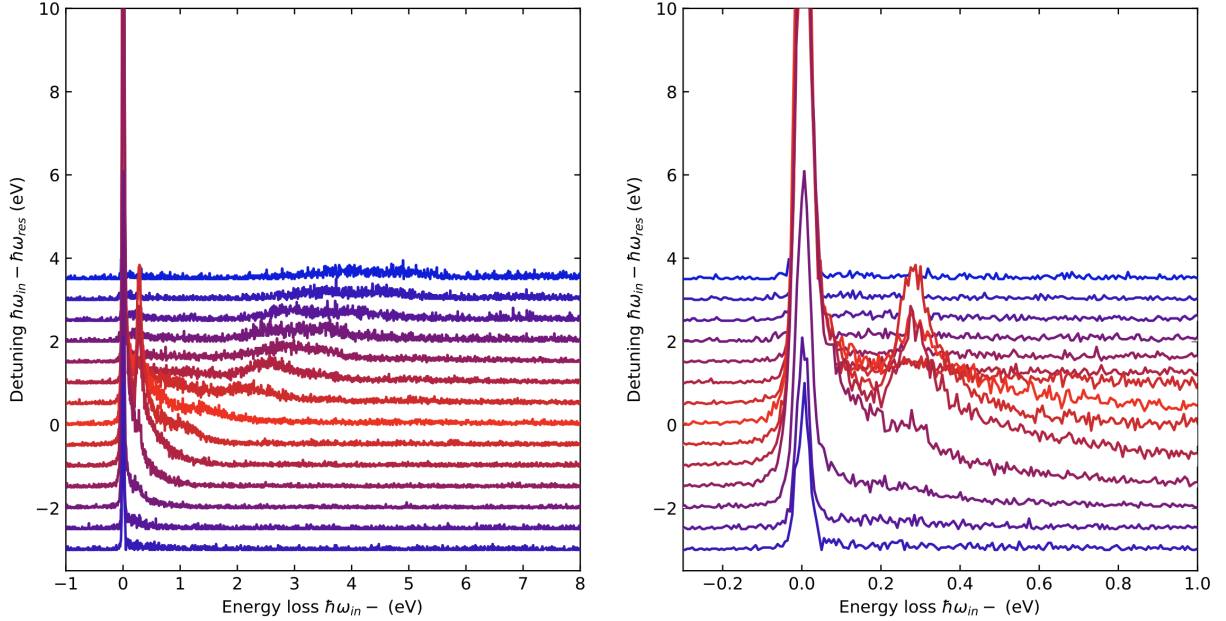

FIG. 7. (a) Waterfall plot illustrating the incident energy dependence of the $M_5$ RIXS response of CePd$_3$. The spectra were obtained at room temperature, in specular scattering geometry at a scattering angle of $2\theta = 150$, with incident x-rays linearly polarised in the scattering plane ($\pi\sigma'+\pi\pi'$). (b) Detailed view of these spectra at low energy transfer.

The fluorescence-like signal is due to excitations of quasiparticles into the highly itinerant conduction band. These Pd $4d$ states are illustrated in the extended DFT+DMFT band structure shown in Supplementary Fig. 8.

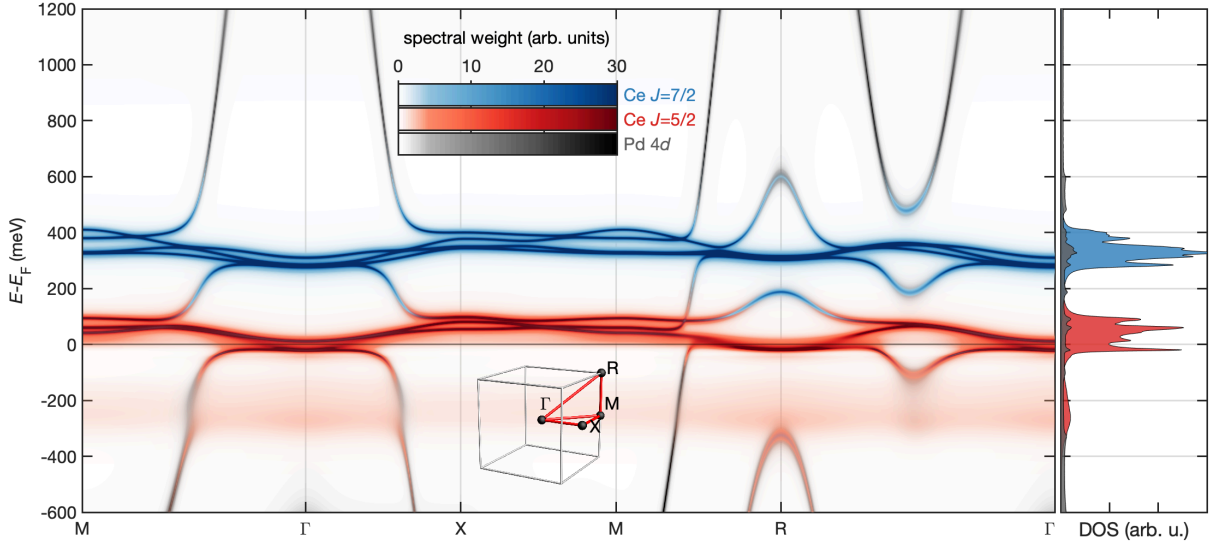

FIG. 8. Extended view of the DFT+DMFT bands at 116 K. The same calculation is shown as in Fig. 1(f) of the manuscript, but over a broader energy range and here including the Pd $4d$ bands. The side panel shows the Ce $J = 7/2$, $J = 5/2$, and Pd $4d$ spectral weight summed along the path shown in the main panel.

10

## Supplementary Note 6: RIXS incident energy and polarization dependence

We combine RIXS polarization-analysis and photon-energy ($\omega_{in}$) dependence to separate the spin-orbit channels of the interband excitations and assign their character. As shown in Supplementary Fig. 9(a), at $\omega_{in} = \omega_{res} - 1.5\,\text{eV}$, the incident energy does not suffice to access intermediate states in which Ce $f^1$ states have been excited to the $J = 7/2$ state. At $\omega_{in} = \omega_{res} - 0.5\,\text{eV}$, as shown in Panel (b), these excitations are strongly enhanced. The polarization analysis of the scattered beam reveals that they appear mostly in the crossed ($\sigma\pi'$) channel. This is further emphasized by subtracting the former from the latter spectra, see Panel (c), middle. The spectral weight attributable to spin-orbit excitations can thus be isolated. In turn, subtracting this signal from the $\omega_{res} - 0.5\,\text{eV}$ data separates excitations within the ground state manifold, see Panel (c), bottom. Overall, both transitions appear with similar spectral weight in the crossed polarization channel. The quasielastic Thompson scattering, as well as a sloping background signal appear in the parallel ($\sigma\sigma'$) channel.

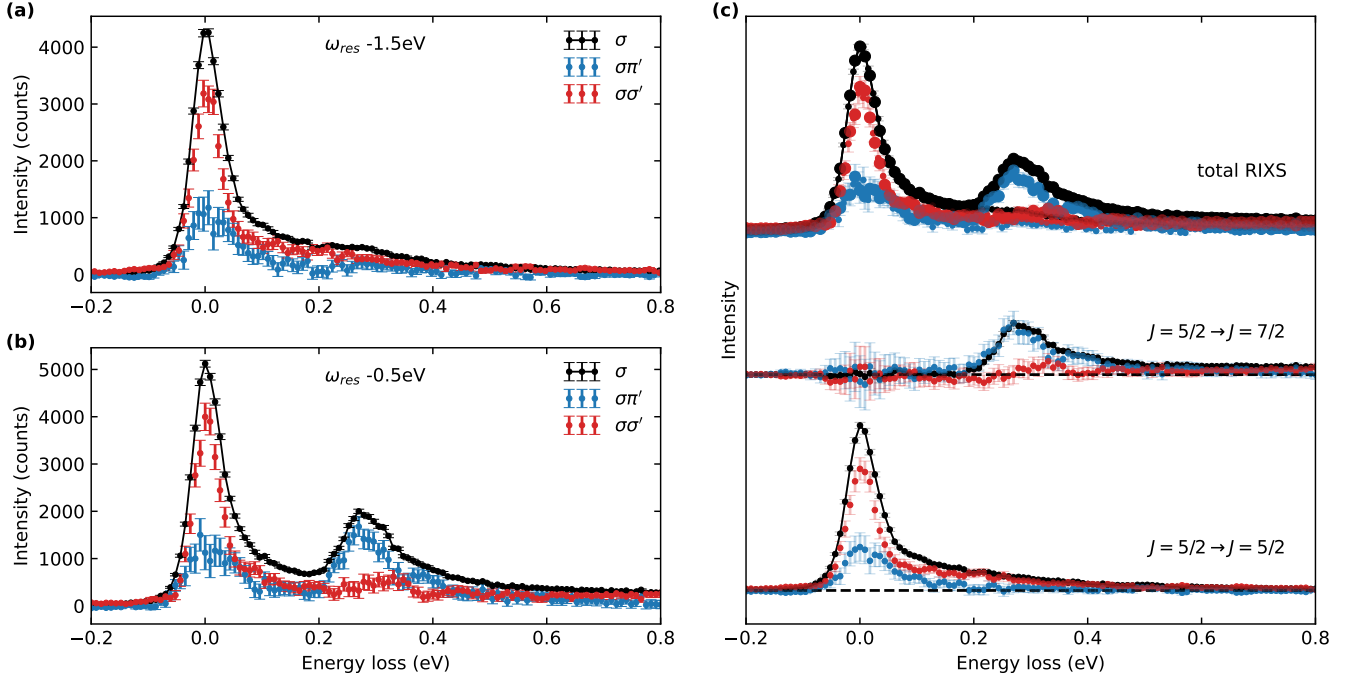

FIG. 9. Polarization and energy dependence of the $M_5$ RIXS response of CePd$_3$. (a) Spectra without polarization analysis, and in the separated $\sigma\pi'$ and $\sigma\sigma'$ channels, at room temperature, 1.5 eV below the resonance ($\approx 880.7\,\text{eV}$). (b) The corresponding spectra obtained 0.5 eV below resonance ($\approx 881.7\,\text{eV}$). (c) Top: Reproduction of the spectra as show in panels (a,b). Middle and bottom: Separation of the spin orbit channels ($J = 5/2 \to 5/2$, vs. $5/2 \to 7/2$), by subtraction of the spectra shown in Panels (a,b).

## Supplementary Note 7: Ce M-edge x-ray absorption

The M-edge x-ray absorption spectroscopy (XAS) characteristics observed in this study are consistent with earlier fluorescence-yield measurements [34] and resemble those of other strongly valence fluctuating Ce materials [35].

In Supplementary Fig. 10, we compare Ce $M$ edge x-ray absorption spectra of CePd$_3$ obtained at the RIXS instrument [in total electron yield (TEY) mode, as shown in Fig. 2(d) of the manuscript] with those obtained in partial-fluorescence-yield (PFY) mode at the MaRES endstation of BL29/BOREAS (ALBA). During the RIXS experiment, TEY-XAS spectra were measured as the drain current from the sample surface (at 22 K). The photon energy range shown here covers both Ce $M_5$ ($3d_{5/2} \to 4f_{7/2}$, 884 eV) and $M_4$ ($3d_{3/2} \to 4f_{5/2}$, 902 eV) absorption edges, split by the spin-orbit coupling of the $3d^9$ core-hole. These TEY-XAS spectra did not vary with the incident angle of the beam, which indicates that the measurement is not strongly affected by surface contamination or reconstruction. The intricate lineshapes reflect the impact of the core-hole potential on the configuration of the largely unoccupied $4f$ manifold, corresponding to the available intermediate states of the RIXS process. The photon energy $h\nu = 882.2$ eV chosen in this study corresponds approximately to the center of the $M_5$ XAS double-peak in trivalent Ce compounds [34, 36, 37].

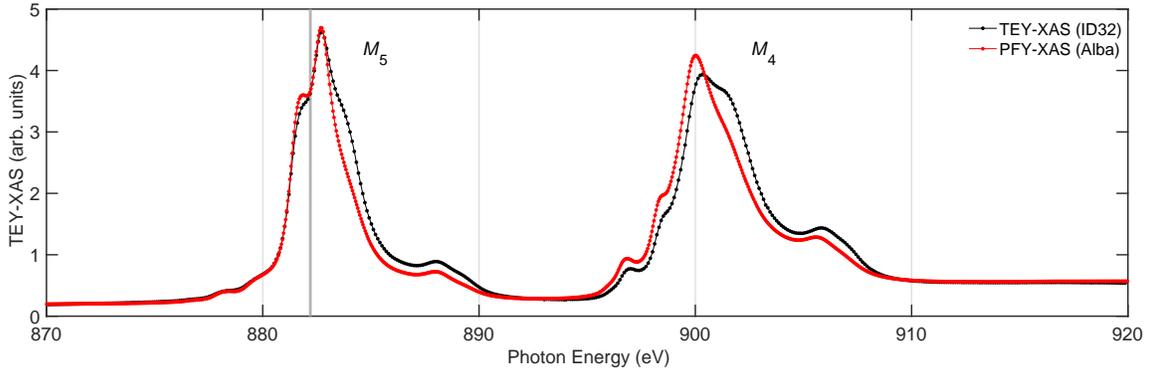

FIG. 10. Cerium $M$ edge x-ray absorption spectra recorded in total electron yield mode at the RIXS instrument (ID32) and in partial fluorescence mode at BL29/BOREAS (ALBA). The resonance energy 882.2 eV, at which the RIXS spectra shown in the manuscript were obtained, is marked by a gray line.

## Supplementary Note 8: Overview of RIXS spectra and fits

Below we provide a complete overview of the temperature-dependent Ce $M_5$ RIXS data discussed in the manuscript. A realistic model of these spectra, taking into account the dispersive $f$-band structure, is currently not feasible. The phenomenological analysis presented in Supplementary Figs. 11 and 12 is therefore not intended as an unequivocal interpretation. For instance, to allow convergence, peak positions have to be fixed in the changeover temperature range around $T_{\text{coh}}$. It is therefore not meaningful to infer the exact positions of individual Lorentzian peaks. The purpose of this analysis is merely to illustrate the changeover between two lineshapes, which is evident even in the raw data.

The spectra at momentum-transfers $\mathbf{Q} = (0.395, 0, 0.395)$ and $\mathbf{Q} = (0.495, 0, 0)$ were each tracked in the temperature range of $22\,\text{K} - 300\,\text{K}$, and are shown in Supplementary Figs. 11 and 12, respectively:

- Panels labeled (a) show the raw data at energy transfers between $-100$ and $700\,\text{meV}$. The (quasi-)elastic scattering is shaded gray. In the $22\,\text{K}$ spectra [see panels (7a)], this contribution can be distinguished as a resolution-limited peak resembling the instrumental resolution function (i.e., a Gaussian distribution of FWHM $32\,\text{meV}$). We attribute this contribution to diffuse elastic scattering (e.g., due to minor bulk and surface structural defects) and hence assume that it is, as inferred from the $22\,\text{K}$ spectra, independent of temperature. Up to four Lorentzian peaks, used as a phenomenological model, are indicated by colored lineshapes. The corresponding fits include a scale parameter $c$, the broadening $\Gamma$, and the resonance energy $E_0$,

$$L(E) = c\,E\,\frac{\Gamma/(2\pi)}{\Gamma^2 + (|E| - E_0)^2} \quad .$$

  Note that the maximum of this function lies at $E_{\text{max}} = \sqrt{E_0^2 + \Gamma^2}$. This lineshape has proven to be a good approximation of the dynamic magnetic susceptibility calculated in NCA/AIM [38, 39] and thus provides an adequate model of the momentum-averaged neutron spectra of many Ce-intermetallics. The Principle of Detailed Balance was applied to these lineshapes to obtain the corresponding anti-Stokes spectral weight without further variables. A $\delta(E)$-like function was used to model the (quasi-)elastic scattering. The sum of these features was convoluted with the experimental resolution function.

- Panels (b) illustrate the same data after subtraction of the (quasi-)elastic scattering and photon-energy–gain intensity. The signal assigned to excitations within the $J = 5/2$ manifold and into the $J = 7/2$ states is indicated by red and blue linehshapes, respectively. The NCA/AIM calculations of $\chi''$ [Parameter Set 1, as shown in Supplementary Fig. 1(a)] are superimposed as black dashed lines.

- Panels (c) show detail views of the very-low energy-transfer regime ($-100$–$50\,\text{meV}$), with an emphasis on photon-energy–gain (anti-Stokes) excitations. The observation of this effect had not been possible with x-ray spectrometers of the previous generation (with energy resolution $\text{d}E > k_{\text{B}}T$). The shaded areas indicate separately the $J = 5/2$ Lorentzian lineshape and the corresponding anti-Stokes intensity, obtained without further parameters by the Principle of Detailed Balance.

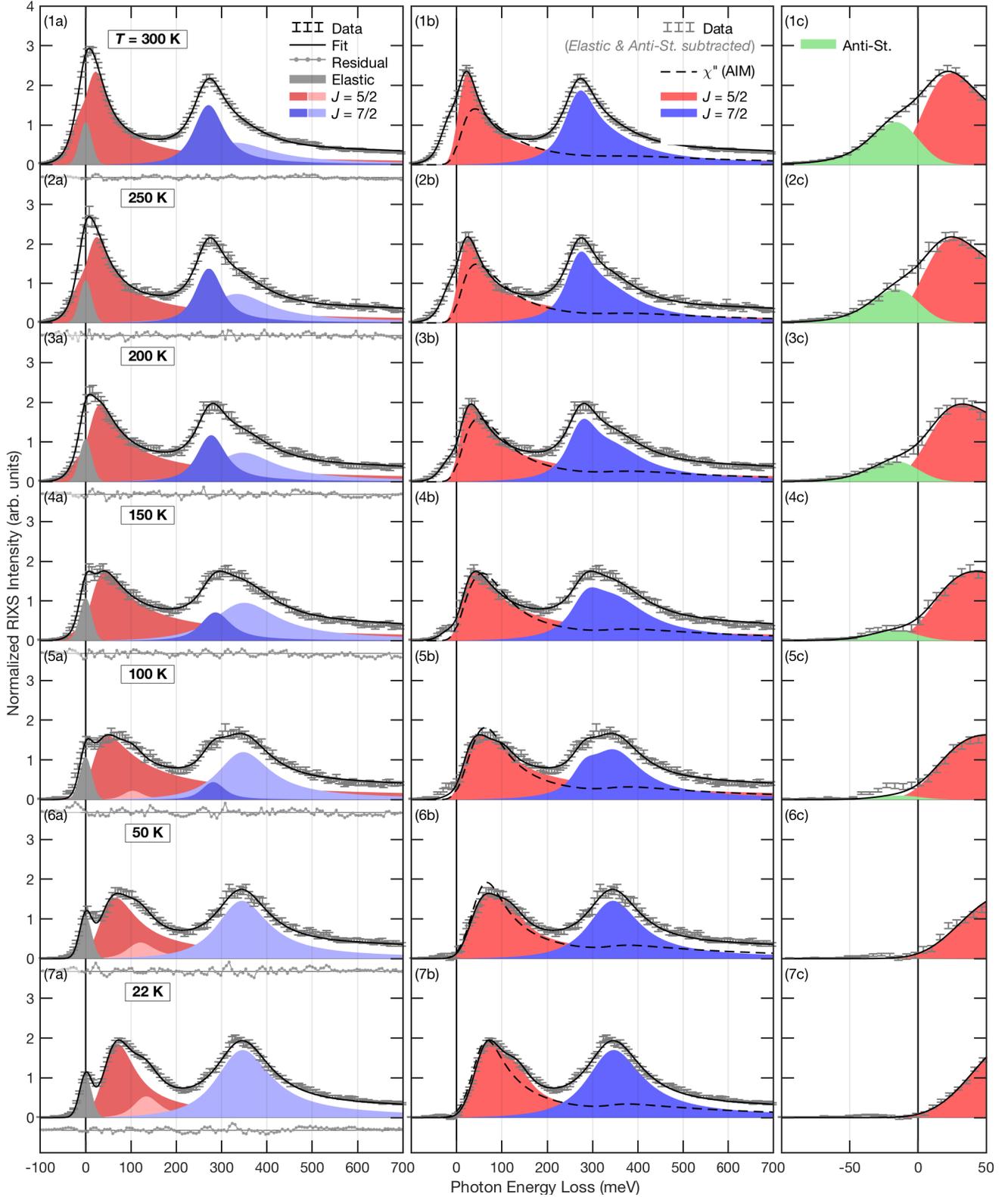

FIG. 11. Overview of CePd$_3$ RIXS spectra measured at momentum transfer $\mathbf{Q} = (0.4, 0, 0.4)$, illustrating the thermal variation in the range of 22 K – 300 K. (1a)–(7a) Raw data and phenomenological fits. Separate Lorentzian lineshapes are indicated by colored lineshapes and the temperature-independent (quasi-)elastic Gaussian peak (32 meV FWHM) is shaded gray. (1b)–(7b) Corresponding view of the spectra after subtraction of the (quasi-)elastic scattering. The NCA/AIM calculations of $\chi''(\omega)$ are shown as back dashed lines. (1c)–(7c) Detail view of the photon-energy–loss regime, emphasizing the evolution of anti-Stokes scattering at high temperatures. The seperated (anti-) Stokes contributions to the low-energy RIXS intensity are indicated as shaded areas.

14

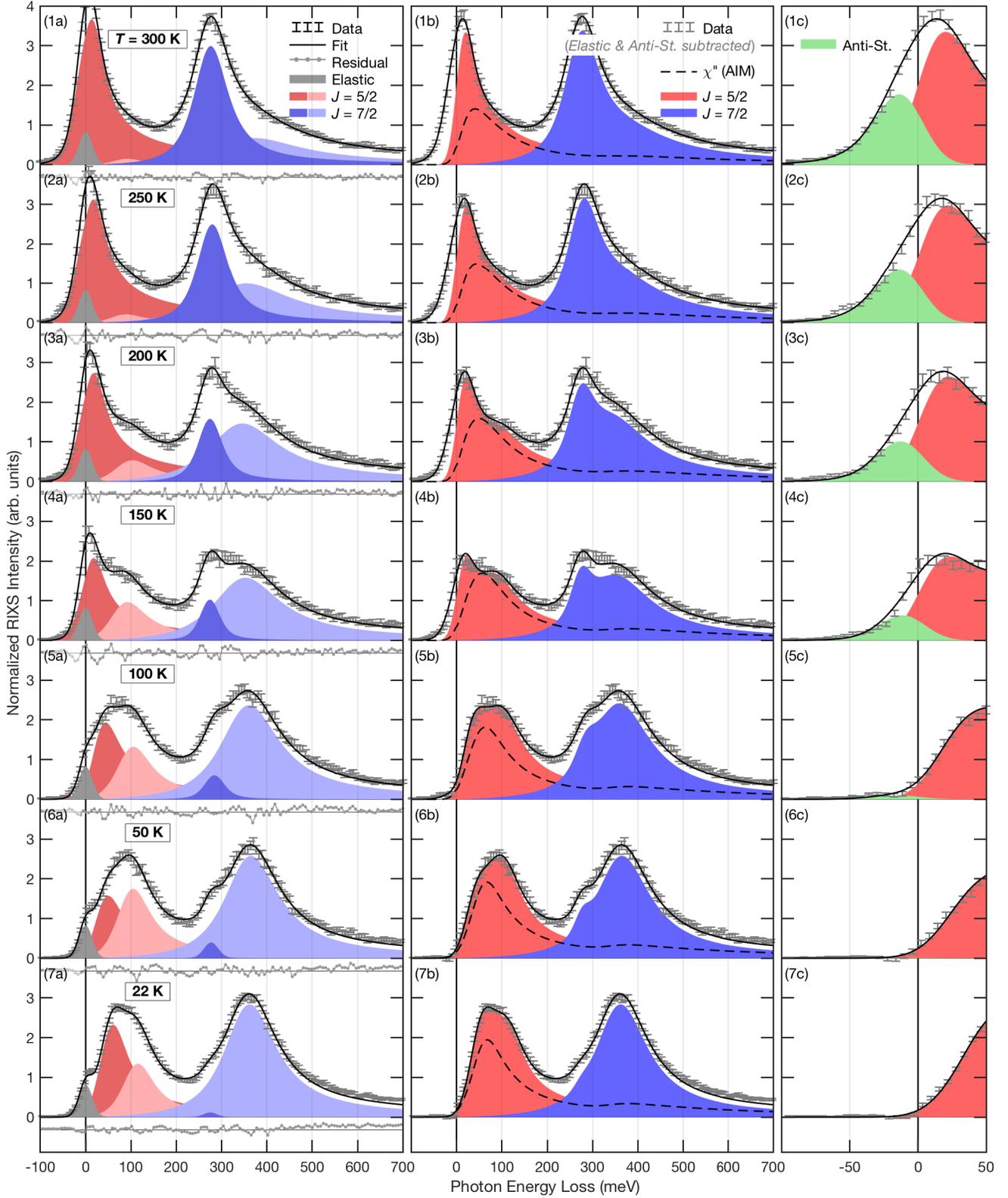

FIG. 12. Overview of CePd$_3$ RIXS spectra measured at a momentum transfer of $\mathbf{Q} = (0.5, 0, 0)$, illustrating the thermal variation in the range of 22 K – 300 K. The data is presented in analogy to Supplementary Fig. 11.

15





\begin{thebibliography}{10}
\expandafter\ifx\csname url\endcsname\relax
  \def\url#1{\texttt{#1}}\fi
\expandafter\ifx\csname urlprefix\endcsname\relax\def\urlprefix{URL }\fi
\providecommand{\bibinfo}[2]{#2}
\providecommand{\eprint}[2][]{\url{#2}}

\bibitem{Paschen2021}
\bibinfo{author}{Paschen, S.} \& \bibinfo{author}{Si, Q.}
\newblock \bibinfo{title}{Quantum phases driven by strong correlations}.
\newblock \emph{\bibinfo{journal}{Nature Reviews Physics}}
  \textbf{\bibinfo{volume}{3}}, \bibinfo{pages}{9–26} (\bibinfo{year}{2021}).
\newblock \eprint{https://doi.org/10.1038/s42254-020-00262-6}.

\bibitem{Shim2007}
\bibinfo{author}{Shim, J.~H.}, \bibinfo{author}{Haule, K.} \&
  \bibinfo{author}{Kotliar, G.}
\newblock \bibinfo{title}{Modeling the localized-to-itinerant electronic
  transition in the heavy fermion system {CeIrIn}$_5$}.
\newblock \emph{\bibinfo{journal}{Science}} \textbf{\bibinfo{volume}{318}},
  \bibinfo{pages}{1615--1617} (\bibinfo{year}{2007}).
\newblock
  \eprint{https://science.sciencemag.org/content/318/5856/1615.full.pdf}.

\bibitem{Haule2010}
\bibinfo{author}{Haule, K.}, \bibinfo{author}{Yee, C.-H.} \&
  \bibinfo{author}{Kim, K.}
\newblock \bibinfo{title}{Dynamical mean-field theory within the full-potential
  methods: Electronic structure of {CeIrIn}$_{5}$, {CeCoIn}$_{5}$, and
  {CeRhIn}$_{5}$}.
\newblock \emph{\bibinfo{journal}{Phys. Rev. B}} \textbf{\bibinfo{volume}{81}},
  \bibinfo{pages}{195107} (\bibinfo{year}{2010}).
\newblock \eprint{https://link.aps.org/doi/10.1103/PhysRevB.81.195107}.

\bibitem{Jang2017}
\bibinfo{author}{Jang, S.} \emph{et~al.}
\newblock \bibinfo{title}{Evolution of the {Kondo} lattice electronic structure
  above the transport coherence temperature}.
\newblock \emph{\bibinfo{journal}{Proceedings of the National Academy of
  Sciences}} \textbf{\bibinfo{volume}{117}}, \bibinfo{pages}{23467--23476}
  (\bibinfo{year}{2020}).
\newblock \eprint{https://www.pnas.org/content/117/38/23467.full.pdf}.

\bibitem{Lawrence2008}
\bibinfo{author}{Lawrence, J.~M.}
\newblock \bibinfo{title}{{Intermediate Valence Metals}}.
\newblock \emph{\bibinfo{journal}{Modern Physics Letters B}}
  \textbf{\bibinfo{volume}{22}}, \bibinfo{pages}{1273--1295}
  (\bibinfo{year}{2008}).
\newblock
  \eprint{http://www.worldscientific.com/doi/abs/10.1142/S0217984908016042}.

\bibitem{Burdin2000}
\bibinfo{author}{Burdin, S.}, \bibinfo{author}{Georges, A.} \&
  \bibinfo{author}{Grempel, D.~R.}
\newblock \bibinfo{title}{Coherence scale of the {Kondo} lattice}.
\newblock \emph{\bibinfo{journal}{Phys. Rev. Lett.}}
  \textbf{\bibinfo{volume}{85}}, \bibinfo{pages}{1048--1051}
  (\bibinfo{year}{2000}).
\newblock \eprint{https://link.aps.org/doi/10.1103/PhysRevLett.85.1048}.

\bibitem{Aynajian2012}
\bibinfo{author}{Aynajian, P.} \emph{et~al.}
\newblock \bibinfo{title}{Visualizing heavy fermions emerging in a quantum
  critical {Kondo} lattice}.
\newblock \emph{\bibinfo{journal}{Nature}} \textbf{\bibinfo{volume}{486}},
  \bibinfo{pages}{201--206} (\bibinfo{year}{2012}).
\newblock \eprint{https://doi.org/10.1038/nature11204}.

\bibitem{Goremychkin2018}
\bibinfo{author}{Goremychkin, E.~A.} \emph{et~al.}
\newblock \bibinfo{title}{{Coherent band excitations in CePd$_3$ : A comparison
  of neutron scattering and ab initio theory}}.
\newblock \emph{\bibinfo{journal}{Science}} \textbf{\bibinfo{volume}{359}},
  \bibinfo{pages}{186--191} (\bibinfo{year}{2018}).
\newblock \eprint{1611.01149}.

\bibitem{Fobes2018}
\bibinfo{author}{Fobes, D.~M.} \emph{et~al.}
\newblock \bibinfo{title}{Tunable emergent heterostructures in a prototypical
  correlated metal}.
\newblock \emph{\bibinfo{journal}{Nature Physics}}
  \textbf{\bibinfo{volume}{14}}, \bibinfo{pages}{456--460}
  (\bibinfo{year}{2018}).
\newblock \eprint{https://doi.org/10.1038/s41567-018-0060-9}.

\bibitem{Pfleiderer2009}
\bibinfo{author}{Pfleiderer, C.}
\newblock \bibinfo{title}{Superconducting phases of $f$-electron compounds}.
\newblock \emph{\bibinfo{journal}{Rev. Mod. Phys.}}
  \textbf{\bibinfo{volume}{81}}, \bibinfo{pages}{1551--1624}
  (\bibinfo{year}{2009}).
\newblock \eprint{https://link.aps.org/doi/10.1103/RevModPhys.81.1551}.

\bibitem{Mydosh2020}
\bibinfo{author}{Mydosh, J.~A.}, \bibinfo{author}{Oppeneer, P.~M.} \&
  \bibinfo{author}{Riseborough, P.~S.}
\newblock \bibinfo{title}{Hidden order and beyond: an
  experimental{\textemdash}theoretical overview of the multifaceted behavior of
  {URu}$_2${Si}$_2$}.
\newblock \emph{\bibinfo{journal}{Journal of Physics: Condensed Matter}}
  \textbf{\bibinfo{volume}{32}}, \bibinfo{pages}{143002}
  (\bibinfo{year}{2020}).
\newblock \eprint{https://doi.org/10.1088/1361-648x/ab5eba}.

\bibitem{Pirie2020}
\bibinfo{author}{Pirie, H.} \emph{et~al.}
\newblock \bibinfo{title}{Imaging emergent heavy {Dirac} fermions of a
  topological {Kondo} insulator}.
\newblock \emph{\bibinfo{journal}{Nature Physics}}
  \textbf{\bibinfo{volume}{16}}, \bibinfo{pages}{52--56}
  (\bibinfo{year}{2020}).
\newblock \eprint{https://doi.org/10.1038/s41567-019-0700-8}.

\bibitem{Kurumaji2019}
\bibinfo{author}{Kurumaji, T.} \emph{et~al.}
\newblock \bibinfo{title}{Skyrmion lattice with a giant topological {Hall}
  effect in a frustrated triangular-lattice magnet}.
\newblock \emph{\bibinfo{journal}{Science}} \textbf{\bibinfo{volume}{365}},
  \bibinfo{pages}{914--918} (\bibinfo{year}{2019}).
\newblock \eprint{https://www.science.org/doi/pdf/10.1126/science.aau0968}.

\bibitem{Jiao2020}
\bibinfo{author}{Jiao, L.} \emph{et~al.}
\newblock \bibinfo{title}{Chiral superconductivity in heavy-fermion metal
  {UTe}$_2$}.
\newblock \emph{\bibinfo{journal}{Nature}} \textbf{\bibinfo{volume}{579}},
  \bibinfo{pages}{523--527} (\bibinfo{year}{2020}).
\newblock \eprint{https://doi.org/10.1038/s41586-020-2122-2}.

\bibitem{Ronning2017}
\bibinfo{author}{Ronning, F.} \emph{et~al.}
\newblock \bibinfo{title}{Electronic in-plane symmetry breaking at field-tuned
  quantum criticality in {CeRhIn}$_5$}.
\newblock \emph{\bibinfo{journal}{Nature}} \textbf{\bibinfo{volume}{548}},
  \bibinfo{pages}{313 EP --} (\bibinfo{year}{2017}).
\newblock \eprint{https://doi.org/10.1038/nature23315}.

\bibitem{Rosa2020}
\bibinfo{author}{Seo, S.} \emph{et~al.}
\newblock \bibinfo{title}{Nematic state in {CeAuSb}$_{2}$}.
\newblock \emph{\bibinfo{journal}{Phys. Rev. X}} \textbf{\bibinfo{volume}{10}},
  \bibinfo{pages}{011035} (\bibinfo{year}{2020}).
\newblock \eprint{https://link.aps.org/doi/10.1103/PhysRevX.10.011035}.

\bibitem{Pagliuso2002}
\bibinfo{author}{Pagliuso, P.} \emph{et~al.}
\newblock \bibinfo{title}{Structurally tuned superconductivity in heavy-fermion
  {Ce$M$In}$_5$ {($M$ = Co, Ir, Rh)}}.
\newblock \emph{\bibinfo{journal}{Physica B: Condensed Matter}}
  \textbf{\bibinfo{volume}{320}}, \bibinfo{pages}{370--375}
  (\bibinfo{year}{2002}).
\newblock
  \eprint{https://www.sciencedirect.com/science/article/pii/S0921452602007512}.

\bibitem{Willers2015}
\bibinfo{author}{Willers, T.} \emph{et~al.}
\newblock \bibinfo{title}{Correlation between ground state and orbital
  anisotropy in heavy fermion materials}.
\newblock \emph{\bibinfo{journal}{Proceedings of the National Academy of
  Sciences}} \textbf{\bibinfo{volume}{112}}, \bibinfo{pages}{2384--2388}
  (\bibinfo{year}{2015}).
\newblock \eprint{https://www.pnas.org/content/112/8/2384.full.pdf}.

\bibitem{Moll2017}
\bibinfo{author}{Moll, P. J.~W.} \emph{et~al.}
\newblock \bibinfo{title}{Emergent magnetic anisotropy in the cubic
  heavy-fermion metal {CeIn}$_3$}.
\newblock \emph{\bibinfo{journal}{npj Quantum Materials}}
  \textbf{\bibinfo{volume}{2}}, \bibinfo{pages}{46} (\bibinfo{year}{2017}).
\newblock \eprint{https://doi.org/10.1038/s41535-017-0052-5}.

\bibitem{Rosa2019}
\bibinfo{author}{Rosa, P. F.~S.} \emph{et~al.}
\newblock \bibinfo{title}{Enhanced hybridization sets the stage for electronic
  nematicity in {CeRhIn}$_{5}$}.
\newblock \emph{\bibinfo{journal}{Phys. Rev. Lett.}}
  \textbf{\bibinfo{volume}{122}}, \bibinfo{pages}{016402}
  (\bibinfo{year}{2019}).
\newblock \eprint{https://link.aps.org/doi/10.1103/PhysRevLett.122.016402}.

\bibitem{Patil2016}
\bibinfo{author}{Patil, S.} \emph{et~al.}
\newblock \bibinfo{title}{Arpes view on surface and bulk hybridization
  phenomena in the antiferromagnetic kondo lattice cerh2si2}.
\newblock \emph{\bibinfo{journal}{Nature Communications}}
  \textbf{\bibinfo{volume}{7}}, \bibinfo{pages}{11029} (\bibinfo{year}{2016}).
\newblock \urlprefix\url{https://doi.org/10.1038/ncomms11029}.

\bibitem{Kummer2015}
\bibinfo{author}{Kummer, K.} \emph{et~al.}
\newblock \bibinfo{title}{{Temperature-independent Fermi surface in the Kondo
  lattice YbRh$_2$Si$_2$}}.
\newblock \emph{\bibinfo{journal}{Physical Review X}}
  \textbf{\bibinfo{volume}{5}}, \bibinfo{pages}{1--9} (\bibinfo{year}{2015}).

\bibitem{Murani1996b}
\bibinfo{author}{Murani, A.~P.}, \bibinfo{author}{Raphel, R.},
  \bibinfo{author}{Bowden, Z.~A.} \& \bibinfo{author}{Eccleston, R.~S.}
\newblock \bibinfo{title}{{Kondo} resonance energies in {CePd}$_3$}.
\newblock \emph{\bibinfo{journal}{Phys. Rev. B}} \textbf{\bibinfo{volume}{53}},
  \bibinfo{pages}{8188--8191} (\bibinfo{year}{1996}).
\newblock \eprint{https://link.aps.org/doi/10.1103/PhysRevB.53.8188}.

\bibitem{Hancock2018}
\bibinfo{author}{Hancock, J.~N.} \emph{et~al.}
\newblock \bibinfo{title}{{Kondo} lattice excitation observed using resonant
  inelastic x-ray scattering at the {Yb} ${M}_{5}$ edge}.
\newblock \emph{\bibinfo{journal}{Phys. Rev. B}} \textbf{\bibinfo{volume}{98}},
  \bibinfo{pages}{075158} (\bibinfo{year}{2018}).
\newblock \eprint{https://link.aps.org/doi/10.1103/PhysRevB.98.075158}.

\bibitem{Amorese2018c}
\bibinfo{author}{Amorese, A.} \emph{et~al.}
\newblock \bibinfo{title}{Crystal electric field in {CeRh}$_2${Si}$_2$ studied
  with high-resolution resonant inelastic soft x-ray scattering}.
\newblock \emph{\bibinfo{journal}{Phys. Rev. B}} \textbf{\bibinfo{volume}{97}},
  \bibinfo{pages}{245130} (\bibinfo{year}{2018}).
\newblock \eprint{https://link.aps.org/doi/10.1103/PhysRevB.97.245130}.

\bibitem{Fanelli2014}
\bibinfo{author}{Fanelli, V.~R.} \emph{et~al.}
\newblock \bibinfo{title}{{Q-dependence of the spin fluctuations in the
  intermediate valence compound CePd3}}.
\newblock \emph{\bibinfo{journal}{J. Phys.: Condens. Matter}}
  \textbf{\bibinfo{volume}{26}}, \bibinfo{pages}{225602}
  (\bibinfo{year}{2014}).

\bibitem{Lawrence1985}
\bibinfo{author}{Lawrence, J.~M.}, \bibinfo{author}{Thompson, J.~D.} \&
  \bibinfo{author}{Chen, Y.~Y.}
\newblock \bibinfo{title}{Two energy scales in {CePd}$_{3}$}.
\newblock \emph{\bibinfo{journal}{Phys. Rev. Lett.}}
  \textbf{\bibinfo{volume}{54}}, \bibinfo{pages}{2537--2540}
  (\bibinfo{year}{1985}).
\newblock \eprint{https://link.aps.org/doi/10.1103/PhysRevLett.54.2537}.

\bibitem{Knafo2003}
\bibinfo{author}{Knafo, W.} \emph{et~al.}
\newblock \bibinfo{title}{Study of low-energy magnetic excitations in
  single-crystalline {CeIn}$_3$ by inelastic neutron scattering}.
\newblock \emph{\bibinfo{journal}{Journal of Physics: Condensed Matter}}
  \textbf{\bibinfo{volume}{15}}, \bibinfo{pages}{3741--3749}
  (\bibinfo{year}{2003}).
\newblock \eprint{https://doi.org/10.1088/0953-8984/15/22/308}.

\bibitem{Thalmeier1984}
\bibinfo{author}{Thalmeier, P.}
\newblock \bibinfo{title}{Bound state of phonons and a crystal field excitation
  in {CeAl}$_2$}.
\newblock \emph{\bibinfo{journal}{Journal of Applied Physics}}
  \textbf{\bibinfo{volume}{55}}, \bibinfo{pages}{1916--1920}
  (\bibinfo{year}{1984}).
\newblock \eprint{https://doi.org/10.1063/1.333518}.

\bibitem{Sundermann2017}
\bibinfo{author}{Sundermann, M.} \emph{et~al.}
\newblock \bibinfo{title}{The quartet ground state in {CeB}$_6$: An inelastic
  x-ray scattering study}.
\newblock \emph{\bibinfo{journal}{{EPL} (Europhysics Letters)}}
  \textbf{\bibinfo{volume}{117}}, \bibinfo{pages}{17003}
  (\bibinfo{year}{2017}).
\newblock \urlprefix\url{https://doi.org/10.1209/0295-5075/117/17003}.

\bibitem{supplemental}
\bibinfo{title}{{Supplemental Information file} attached to this preprint}.

\bibitem{Murani2002b}
\bibinfo{author}{Murani, A.~P.}, \bibinfo{author}{Reske, J.},
  \bibinfo{author}{Ivanov, A.~S.} \& \bibinfo{author}{Palleau, P.}
\newblock \bibinfo{title}{{Evolution of the spin-orbit excitation across the
  gamma-alpha transition in Ce}}.
\newblock \emph{\bibinfo{journal}{Phys. Rev. B}} \textbf{\bibinfo{volume}{65}},
  \bibinfo{pages}{094416} (\bibinfo{year}{2002}).

\bibitem{Dvorak2016}
\bibinfo{author}{Dvorak, J.}, \bibinfo{author}{Jarrige, I.},
  \bibinfo{author}{Bisogni, V.}, \bibinfo{author}{Coburn, S.} \&
  \bibinfo{author}{Leonhardt, W.}
\newblock \bibinfo{title}{{Towards 10 meV resolution: The design of an
  ultrahigh resolution soft X-ray RIXS spectrometer}}.
\newblock \emph{\bibinfo{journal}{Review of Scientific Instruments}}
  \textbf{\bibinfo{volume}{87}}, \bibinfo{pages}{115109}
  (\bibinfo{year}{2016}).
\newblock \eprint{http://aip.scitation.org/doi/10.1063/1.4964847}.

\bibitem{Brookes2018}
\bibinfo{author}{Brookes, N.} \emph{et~al.}
\newblock \bibinfo{title}{The beamline {ID32} at the {ESRF} for soft x-ray high
  energy resolution resonant inelastic x-ray scattering and polarisation
  dependent x-ray absorption spectroscopy}.
\newblock \emph{\bibinfo{journal}{Nuclear Instruments and Methods in Physics
  Research Section A: Accelerators, Spectrometers, Detectors and Associated
  Equipment}} \textbf{\bibinfo{volume}{903}}, \bibinfo{pages}{175 -- 192}
  (\bibinfo{year}{2018}).
\newblock
  \eprint{http://www.sciencedirect.com/science/article/pii/S0168900218308234}.

\bibitem{Kummer2017}
\bibinfo{author}{Kummer, K.} \emph{et~al.}
\newblock \bibinfo{title}{{{\it RixsToolBox}: software for the analysis of soft
  X-ray RIXS data acquired with 2D detectors}}.
\newblock \emph{\bibinfo{journal}{Journal of Synchrotron Radiation}}
  \textbf{\bibinfo{volume}{24}}, \bibinfo{pages}{531--536}
  (\bibinfo{year}{2017}).
\newblock \eprint{https://doi.org/10.1107/S1600577517000832}.

\bibitem{Braicovich2014}
\bibinfo{author}{Braicovich, L.} \emph{et~al.}
\newblock \bibinfo{title}{The simultaneous measurement of energy and linear
  polarization of the scattered radiation in resonant inelastic soft x-ray
  scattering}.
\newblock \emph{\bibinfo{journal}{Review of Scientific Instruments}}
  \textbf{\bibinfo{volume}{85}}, \bibinfo{pages}{115104}
  (\bibinfo{year}{2014}).
\newblock \eprint{https://doi.org/10.1063/1.4900959}.

\bibitem{Bickers1987b}
\bibinfo{author}{Bickers, N.~E.}
\newblock \bibinfo{title}{{Review of techniques in the large-N expansion for
  dilute magnetic alloys}}.
\newblock \emph{\bibinfo{journal}{Rev. Mod. Phys.}}
  \textbf{\bibinfo{volume}{59}}, \bibinfo{pages}{845} (\bibinfo{year}{1987}).

\bibitem{Lawrence2001}
\bibinfo{author}{Lawrence, J.~M.} \emph{et~al.}
\newblock \bibinfo{title}{{Slow crossover in Yb$X$Cu$_4$ ($X=$ Ag, Cd, In, Mg,
  Tl, Zn) intermediate-valence compounds}}.
\newblock \emph{\bibinfo{journal}{Physical Review B}}
  \textbf{\bibinfo{volume}{63}}, \bibinfo{pages}{054427}
  (\bibinfo{year}{2001}).
\newblock \eprint{https://link.aps.org/doi/10.1103/PhysRevB.63.054427}.

\bibitem{Blaha2018}
 \emph{\bibinfo{title}{{An Augmented Plane Wave + Local Orbitals Program for
  Calculating Crystal Properties}}}.
\newblock \bibinfo{number}{{ISBN 3-9501031-1-2}}
  (\bibinfo{publisher}{{Karlheinz Schwarz, Techn. Universit{\"a}t Wien,
  Austria}}, \bibinfo{year}{2018}).

\bibitem{Blaha2020}
\bibinfo{author}{Blaha, P.} \emph{et~al.}
\newblock \bibinfo{title}{{WIEN2k}: An {APW+lo} program for calculating the
  properties of solids}.
\newblock \emph{\bibinfo{journal}{The Journal of Chemical Physics}}
  \textbf{\bibinfo{volume}{152}}, \bibinfo{pages}{074101}
  (\bibinfo{year}{2020}).
\newblock \eprint{https://doi.org/10.1063/1.5143061}.

\bibitem{Werner2006}
\bibinfo{author}{Werner, P.}, \bibinfo{author}{Comanac, A.},
  \bibinfo{author}{de' Medici, L.}, \bibinfo{author}{Troyer, M.} \&
  \bibinfo{author}{Millis, A.~J.}
\newblock \bibinfo{title}{Continuous-time solver for quantum impurity models}.
\newblock \emph{\bibinfo{journal}{Phys. Rev. Lett.}}
  \textbf{\bibinfo{volume}{97}}, \bibinfo{pages}{076405}
  (\bibinfo{year}{2006}).
\newblock \eprint{https://link.aps.org/doi/10.1103/PhysRevLett.97.076405}.

\bibitem{Boehnke2011}
\bibinfo{author}{Boehnke, L.}, \bibinfo{author}{Hafermann, H.},
  \bibinfo{author}{Ferrero, M.}, \bibinfo{author}{Lechermann, F.} \&
  \bibinfo{author}{Parcollet, O.}
\newblock \bibinfo{title}{Orthogonal polynomial representation of
  imaginary-time {Green's} functions}.
\newblock \emph{\bibinfo{journal}{Phys. Rev. B}} \textbf{\bibinfo{volume}{84}},
  \bibinfo{pages}{075145} (\bibinfo{year}{2011}).
\newblock \eprint{https://link.aps.org/doi/10.1103/PhysRevB.84.075145}.

\bibitem{Hafermann2012}
\bibinfo{author}{Hafermann, H.}, \bibinfo{author}{Patton, K.~R.} \&
  \bibinfo{author}{Werner, P.}
\newblock \bibinfo{title}{Improved estimators for the self-energy and vertex
  function in hybridization-expansion continuous-time quantum {Monte Carlo}
  simulations}.
\newblock \emph{\bibinfo{journal}{Phys. Rev. B}} \textbf{\bibinfo{volume}{85}},
  \bibinfo{pages}{205106} (\bibinfo{year}{2012}).
\newblock \eprint{https://link.aps.org/doi/10.1103/PhysRevB.85.205106}.

\bibitem{Hariki2015}
\bibinfo{author}{Hariki, A.}, \bibinfo{author}{Yamanaka, A.} \&
  \bibinfo{author}{Uozumi, T.}
\newblock \bibinfo{title}{Theory of spin-state selective nonlocal screening in
  {Co} $2p$ x-ray photoemission spectrum of {LaCoO}$_3$}.
\newblock \emph{\bibinfo{journal}{Journal of the Physical Society of Japan}}
  \textbf{\bibinfo{volume}{84}}, \bibinfo{pages}{073706}
  (\bibinfo{year}{2015}).
\newblock \eprint{https://doi.org/10.7566/JPSJ.84.073706}.

\bibitem{Malterre1992}
\bibinfo{author}{Malterre, D.}, \bibinfo{author}{Grioni, M.},
  \bibinfo{author}{Weibel, P.}, \bibinfo{author}{Dardel, B.} \&
  \bibinfo{author}{Baer, Y.}
\newblock \bibinfo{title}{Evidence of a {Kondo} scale from the temperature
  dependence of inverse photoemission spectroscopy of {CePd}$_3$}.
\newblock \emph{\bibinfo{journal}{Phys. Rev. Lett.}}
  \textbf{\bibinfo{volume}{68}}, \bibinfo{pages}{2656--2659}
  (\bibinfo{year}{1992}).
\newblock \eprint{https://link.aps.org/doi/10.1103/PhysRevLett.68.2656}.

\bibitem{Souma2001}
\bibinfo{author}{Souma, S.}, \bibinfo{author}{Kumigashira, H.},
  \bibinfo{author}{Ito, T.}, \bibinfo{author}{Takahashi, T.} \&
  \bibinfo{author}{Kasaya, M.}
\newblock \bibinfo{title}{Ultrahigh-resolution photoemission study of
  {CePd}$_3$: absence of {Kondo} insulator gap}.
\newblock \emph{\bibinfo{journal}{Journal of Electron Spectroscopy and Related
  Phenomena}} \textbf{\bibinfo{volume}{114-116}}, \bibinfo{pages}{735 -- 740}
  (\bibinfo{year}{2001}).
\newblock
  \eprint{http://www.sciencedirect.com/science/article/pii/S0368204800003868}.

\bibitem{Hariki2020}
\bibinfo{author}{Hariki, A.}, \bibinfo{author}{Winder, M.},
  \bibinfo{author}{Uozumi, T.} \& \bibinfo{author}{Kune\ifmmode~\check{s}\else
  \v{s}\fi{}, J.}
\newblock \bibinfo{title}{$\mathrm{LDA}+\mathrm{DMFT}$ approach to resonant
  inelastic x-ray scattering in correlated materials}.
\newblock \emph{\bibinfo{journal}{Phys. Rev. B}}
  \textbf{\bibinfo{volume}{101}}, \bibinfo{pages}{115130}
  (\bibinfo{year}{2020}).
\newblock \eprint{https://link.aps.org/doi/10.1103/PhysRevB.101.115130}.

\end{thebibliography}
\end{document}